\documentclass[11pt,twoside]{article}
\usepackage{amsmath,amssymb}
\usepackage{algorithm}
\usepackage{color,multirow,graphicx}
\RequirePackage{hyperref, natbib}
\RequirePackage{hypernat}

\newtheorem{proposition}{Proposition}
\newtheorem{theorem}{Theorem}
\newtheorem{proof}{Proof}
\newtheorem{corollary}{Corollary}

\DeclareMathOperator{\X}{\mathbf{X}}
\DeclareMathOperator{\Omeg}{\mathbf{\Omega}}

\DeclareMathOperator{\Sig}{\mathbf{\Sigma}}
\DeclareMathOperator{\Sigi}{\mathbf{\Sigma}^{-1}}
\DeclareMathOperator{\Delt}{\mathbf{\Delta}}
\DeclareMathOperator{\Delti}{\mathbf{\Delta}^{-1}}
\DeclareMathOperator{\U}{\mathbf{U}}
\DeclareMathOperator{\D}{\mathbf{D}}

\DeclareMathOperator{\B}{\mathbf{B}}

\DeclareMathOperator{\V}{\mathbf{V}}
\DeclareMathOperator{\Z}{\mathbf{Z}}

\DeclareMathOperator{\M}{\mathbf{M}}
\DeclareMathOperator{\Smat}{\mathbf{S}}
\DeclareMathOperator{\N}{\mathbf{N}}

\DeclareMathOperator{\x}{\mathbf{x}}
\DeclareMathOperator*{\eqdist}{=}

\DeclareMathOperator*{\simiid}{\sim}

\begin{document}

\title{Inference with
  Transposable Data:  Modeling the 
  Effects of Row and Column Correlations}
\author{Genevera I. Allen \\ \& \\ Robert Tibshirani}
% \address{Department of Statistics,
% Stanford University
% Stanford, CA, USA.}
%\email{giallen@stanford.edu}

% \address{Departments of Health Research \& Policy and Statistics,
% Stanford University
% Stanford, CA, USA.}
%\email{tibs@stanford.edu}

\maketitle

\begin{abstract}
We consider the problem of large-scale inference on the row or column
variables of data in the form of a matrix.  Often this data is {\em
  transposable}, 
meaning that both the row variables and column variables are of
potential interest.  An  
example of this scenario is detecting significant genes in microarrays
when the samples or arrays may be dependent due to underlying
relationships.
We study the effect of both row and column
correlations on commonly used test-statistics, null distributions, and multiple
testing procedures, by explicitly modeling the covariances
with the matrix-variate normal distribution.  Using this model, we
give both theoretical and simulation results revealing the problems
associated with
using standard statistical methodology on transposable data.
We solve these problems by estimating the row
and column covariances simultaneously, with transposable regularized
covariance models, and de-correlating or {\em sphering} the
data as a pre-processing step.  Under reasonable assumptions,
our method gives test 
statistics that follow the scaled theoretical null distribution and are
approximately independent.  
Simulations based on various models with structured and observed covariances
from real microarray 
data reveal that our method offers substantial improvements in two
areas: 1) increased statistical power 
and 2) correct estimation of false discovery rates.    \\
\\
{\bf Keywords:} { \em multiple testing, false discovery
  rate, transposable regularized covariance models,
  large-scale inference, covariance estimation, matrix-variate normal,
empirical null}
\end{abstract}

\section{Introduction}

As statisticians, we often make assumptions when constructing a model to
ease computations or employ existing methodologies.  When analyzing
matrix data, we often assume that the variables along one dimension
(say the columns) are independent, allowing us to pool these
observations to make inferences on the variables along the other
dimension (rows).  In microarrays, for example, it is common to
assume that the arrays are independent observations when computing
test statistics, allowing us to assess differential expression in
genes.  Since we are testing many
row variables (for example, genes) simultaneously, we commonly correct
for multiple testing using procedures that theoretically are known
only to control error measures
when the row variables are independent or follow limited dependence
structures.  Thus, for inference with matrix data, we often make
assumptions of independence or limited dependencies among the row
variables and among the column variables to be able to employ existing
statistical methodologies.  What if these assumptions are incorrect?
What if this matrix data is in fact {\em transposable}, meaning that
potentially 
both the rows and/or columns are correlated?  

In this paper, we consider the problem of testing the
significance of row variables in a data matrix where there are
correlations among the rows, or among the
columns, or among both.  We study the behavior of standard statistical
methodology on transposable data and then 
propose a method to directly account for the dependencies when
conducting inference.

Throughout this paper, we often refer to the example of detecting
genes that are differentially expressed between two classes in
microarray data.  
These genomic datasets contain complicated
correlation structures.  Genes in similar pathways, for example, are
usually highly positively correlated.  Other genes may encode
proteins that  
act as inhibitors leading to negative
correlations.  In 
the analysis of microarrays, it is common to assume that the arrays
are independent.  Many have suggested, however, that this may not be
correct \citep{owen_2005, qiu_2005, leek_2008, efron_2009_indep}, due
to the measurement process or latent 
variables.  Arranged in the form of a matrix, this means that both the
row (gene) and column (array) variables could be dependent, indicating
that the data could be {\em transposable}.

While we focus on the example of detecting significant
genes in the two-class microarray, our methods can be
applied to many examples of large-scale inference with transposable
data.  These include: testing the significance of proteins, genes, or
isoforms in data such as protein arrays and next-generation sequencing
data, testing the significance of voxels in functional magnetic
resonance imaging data, and testing the significance of biomarkers in
three-way data where measurements are taken on multiple subjects at
several time points or in many different laboratories.   In all of
these examples, the assumptions of independence along one dimension of
the data is questionable.

\begin{figure}[!!t]
\begin{center}
\includegraphics[width=5in]{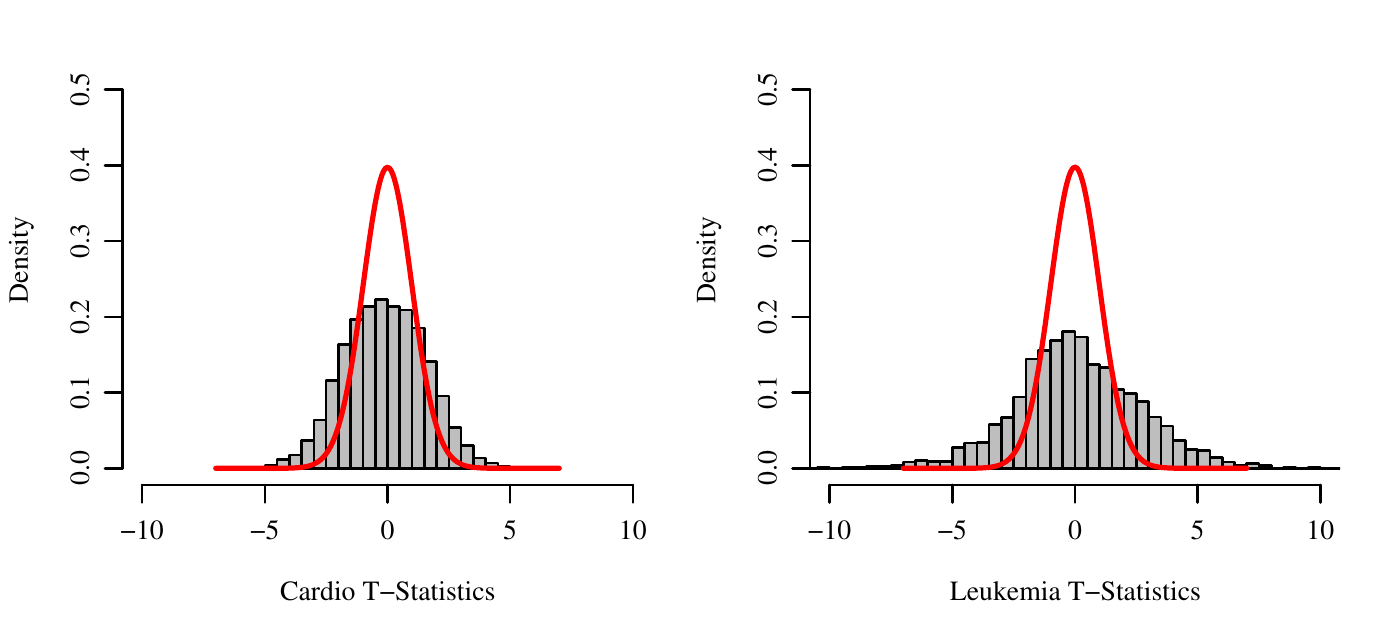}
\end{center}
\caption{ \em  \footnotesize Histograms of two-sample $t$-statistics
  for the ``Cardio'' data 
  (left) and the ``Leukemia'' data (right).  Log intensity values were
  used with the genes and arrays centered.  The theoretical null
  distribution, the 
$t$-distribution with 70 degrees of freedom (left) and 61 degrees of
freedom (right), is drawn in red.  \label{fig_tstat_micro}
}
\end{figure}

We begin by introducing two examples that we will refer to
throughout this paper.  The first is a two-class
microarray study of cardiovascular disease \citep{efron_2009_indep}.
We will refer to this as 
the ``Cardio'' data.  This data has $m = 20,426$ genes and $n = 63$
arrays consisting of 44 controls and 19 diseased patients.  The second
is a two-class microarray study of two types of Leukemia cancer
\citep{golub}, which we will refer to as the ``Leukemia'' data.  This
data has $m = 3,701$ filtered genes and $n = 72$ arrays with 25 and 47
samples in each subtype.  For each of these datasets, we calculate the
two-sample $t$-statistic for each gene and compare their distribution
to that of the theoretical null distribution in Figure
\ref{fig_tstat_micro}.  We 
see that the $t$-statistics are over-dispersed compared to their 
theoretical null distributions.  This could be due to the highly
correlated nature 
of the thousands of genes, or another cause could be correlations
among the arrays.  In fact, the permutation tests of
\citet{efron_2009_indep} reject the null hypothesis of independent
arrays for both of these microarrays.

When studying inference with transposable data, the effects of row and
column correlations must be considered separately.  Since the columns
are generally considered to be independent, population column
correlations lead to the used of incorrect test statistics and null
distributions which in turn result in problems when correcting for
multiple testing.  Row correlations lead to the much discussed problem
of multiple testing dependence \citep{hommel_1986, benjamini_2001,
  storey_2004, leek_2008, sarkar_2008}.

We propose to study and solve these problems by modeling row and column
correlations using the 
mean-restricted matrix-variate normal distribution \citep{allen_2010}
described in Section \ref{section_theory}.  The first half of our paper
is devoted to studying the effects of these correlations on test
statistics and 
their theoretical null distributions, Section \ref{section_nulls},
and on power and multiple testing procedures through a simulation
study in Section \ref{section_sim_study}.  Interestingly, this study
finds the following results.
\begin{enumerate}
\item Unanticipated column correlations dramatically alter the null
  distributions of 
  test statistics leading to the use of incorrect test statistics,
  null distributions and estimates of the FDR.
\item Row correlations do not seem to affect the estimates of the FDR.
\end{enumerate}

The later half of our paper
is focused on solving the problems associated with row and column
correlations by directly making use of the correlation structure.  In
Section \ref{section_de_cor}, we 
simultaneously estimate row and column covariances using transposable
regularized covariance models \citep{allen_2010}.  We then present an
algorithm to {\em sphere} or de-correlate the rows and columns so that
they are approximately independent.  This algorithm is to be used as a
pre-processing step and in
conjunction with standard multiple testing procedures.  Simulation
results using our sphering 
algorithm are presented in Section \ref{section_res} under various
models on both
structured covariance and real microarray covariance examples.  These
reveal two important results:
\begin{enumerate}
\item[(c)] Sphering can alter the rank of the test statistics leading
  to an ordering with higher 
  statistical power.
\item[(d)] Sphering often leads to substantial improvements in the
  estimation of the FDR. 
\end{enumerate}
We conclude with a 
discussion of our study and methods in Section \ref{section_discussion}.

\section{Theoretical Framework}
\label{section_theory}

In this section, we first present a matrix decomposition model
based on the 
mean-restricted matrix-variate normal in Section
\ref{section_model}.  Then, going back to the two-class microarray
example, we consider the test statistic for a
single gene.  Since the arrays are usually assumed to be independent,
the two-sample $z$ and $t$-tests are used commonly to assess differential
expression.  We give the theoretical null
distributions for these test statistics under our
model with column correlations in Section \ref{section_nulls}. 

\subsection{Model}
\label{section_model}

We propose to study row and column correlations through a simple
matrix decomposition model based on the matrix-variate normal.
We motivate the use of this distribution through the example of
microarrays.  

In microarray data, the genes are often assumed to
follow a multivariate normal distribution with the arrays 
independent and identically distributed.  Since we aim to study the
effects of array correlations, we need a parametric model that has the
flexibility to model either array independence or various array
correlation structures.  To this end, we turn to the mean-restricted
matrix-variate normal introduced in \citet{allen_2010}.  (We also
note that \citet{efron_2009_indep} proposes the matrix-variate normal as a
model for microarrays).  This distribution, denoted as $\X \sim
N_{m,n}( \nu, \mu, \Sig, \Delt)$, has separate mean and covariance
parameters for the rows, $\nu \in \Re^{m}$ and $\Sig \in \Re^{m \times
  m}$, and columns, $\mu \in \Re^{n}$ and
$\Delt \in \Re^{n \times n}$.  Thus, we can model array correlations
directly though the covariance matrix $\Delt$.  If the matrix is
transformed into a vector of length $np$, we have that
$\mathrm{vec}(\X) \sim N( \mathrm{vec}(\M), \Omeg)$, where $\M = \nu
\mathbf{1}^{T}_{(n)} + \mathbf{1}_{(m)}$ and $\Omeg = \Delt \otimes \Sig$.
Also, the commonly
used multivariate normal is a special case of the distribution.  If
$\Delt = \mathbf{I}$ and $\mu = \mathbf{0}$, then $\X \sim N(\nu,
\Sig)$.  In fact, all marginal models of the matrix-variate normal are
multivariate normal, meaning that both the genes and arrays separately
are multivariate normal.  Further properties of this distribution are
given in \citet{allen_2010}.

In our matrix decomposition model, we will assume that the data,
$\X$, has $m$ rows and $n$ columns.   We define the overall
row means as $\nu \in \Re^{m}$ and column means as $\mu \in \Re^{n}$.
The covariance of the rows is $\Sig \in \Re^{m \times m}$ and the
covariance of the columns is $\Delt \in \Re^{n \times n}$. 
Then, we decompose the data into a mean, signal, and correlated noise matrix as
follows. 
\begin{align}
\label{decomp}
 \X_{m \times n} &= \M_{m \times n} +  \Smat_{m \times n} + \N_{m \times
  n}, 
\end{align}
\vspace{-8mm}
\begin{align*}
  \textrm{where  }  \M &= \nu \mathbf{1}^{T}_{(n)} + \mathbf{1}_{(m)}
\mu^{T} \textrm{  (mean matrix)},  \nonumber \\
  \Smat & \textrm{ is problem specific}   \textrm{  (signal
  matrix)}, \nonumber \\
 \N &\sim N_{m,n}( \mathbf{0}, \mathbf{0}, \Sig, \Delt) \textrm{ (noise
 matrix)}. \nonumber 
\end{align*}
Thus, $\X - \Smat \sim N_{m,n}( \nu, \mu, \Sig, \Delt)$, meaning that
after removing the signal, the data follows a 
mean-restricted matrix-variate normal distribution.  

For the example of the two-class microarray, we let there be $n_{1}$
arrays in class one, 
with indices denoted by $\mathcal{C}_{1}$, 
and $n_{2}$ in class two, $\mathcal{C}_{2}$.  (For simplicity of
notation, we assume that the first $n_{1}$ arrays are in class one and
the last $n_{2}$ arrays are in class two.)
The class signals, or the gene means for each class are defined as
$\psi_{1} \in \Re^{m}$ and $\psi_{2} \in \Re^{m}$.  Then, the signal
matrix, $\Smat$, can be written as follows.
\begin{align*}
\Smat = \left[ \psi_{1} \mathbf{1}^{T}_{(n_{1})} \hspace{2mm}
  \psi_{2} \mathbf{1}^{T}_{(n_{2})}  \right].
\end{align*}

There are several remarks to make regarding this model.  First, prior
to analyzing data, it is common to standardize the rows. Sometimes
this two-way data is {\em doubly-standardized}, or both the rows and
columns are iteratively scaled \citep{efron_2009_indep, olshen_2010}.  Here, we
center both the rows and columns through the mean matrix $\M$, but do
not directly scale them.  Instead, we allow the
diagonals of the covariance matrices of the rows, $\Sig$, and columns
$\Delt$, to capture the differences in variablities.  Thus, our model
keeps the mean and variances separate in the
estimation process.

\subsection{Null Distributions: The Two-Class Problem}
\label{section_nulls}

In this section, we study the effect of column correlations on the
theoretical null distribution of two-sample test statistics computed for a
single row of the data matrix.   More specifically, we calculate the
distributions of test statistics under our matrix decomposition model
instead of the typical two-sample framework where samples are drawn
independently from two populations.  This corresponds to considering a 
single test for differential expression of gene $i$ between the two
classes.

In the familiar two-sample hypothesis testing problem, we have
a vector $\x = [ \x_{1} \hspace{2mm} \x_{2} ]$ with $\x_{1}$ of length
$n_{1}$ and $\x_{2}$ of length $n_{2}$ where the elements of each
vector are
$x_{1,i} \displaystyle\simiid^{iid}  N( \psi_{1} , \sigma^{2} )$  and $x_{2,i}
\displaystyle\simiid^{iid} N (\psi_{2}, 
\sigma^{2})$.  We wish to test whether there is a
shift in means 
between the two classes, namely 
\begin{align}
\label{hypoth}
H_{0}: \hspace{2mm} \psi_{1} = \psi_{2} \hspace{2mm} \mathrm{vs.}
\hspace{2mm} H_{1}: \hspace{2mm} \psi_{1} \neq \psi_{2}.
\end{align} 
Throughout this paper, we will assume that the variances $\sigma^{2}$
are equal between the two classes, a common assumption in microarrays.

If the variance, $\sigma^{2}$ is known, we have the familiar
two-sample $Z$-statistic,
\begin{align*}
Z = \displaystyle\frac{ \bar{x}_{1} - \bar{x}_{2} }{ \sigma
  \sqrt{c_{n} }}, \textrm{  with } Z \sim N \left(
  \frac{\psi_{1} - \psi_{2}}{\sigma \sqrt{ c_{n} } } , 1 \right),
\end{align*}
where $\bar{x}_{k} = \frac{1}{n_{k}} \sum_{i=1}^{n_{k}}
x_{i}$ and $c_{n} = \frac{1}{n_{1}} + \frac{1}{n_{2}}$.
 
Now, going back to our matrix decomposition model, we wish to know
the distribution of the $Z$-statistic for each row when there are column
correlations.
\begin{theorem}
\label{prop1}
Let $\x = [
\x_{1} \hspace{2mm} 
\x_{2} ] \sim N_{1,n} \left( 0, [ \psi_{1} \mathbf{1}_{(n_{1})}
  \hspace{2mm} \psi_{2} \mathbf{1}_{(n_{2})} ],
  \sigma^{2}, \Delt \right)$.  Then,
\begin{align}
Z \sim N \left( \frac{ \psi_{1} - \psi_{2}  }{ \sigma
  \sqrt{c_{n} }} , \frac{ \eta}{ c_{n}} \right) 
\end{align}
$\textrm{  where } \eta
\triangleq \displaystyle\sum_{j=1}^{n} \left( \frac{1}{n_{1}} \sum_{i \in
    \mathcal{C}_{1}} L_{ij} - \frac{1}{n_{2}} \sum_{i \in
    \mathcal{C}_{2}} L_{ij}  \right)^{2}$ and 
 $\mathbf{L}$ is the matrix square root of $\Delt$.
\end{theorem}

In terms of the decomposition
\eqref{decomp}, the assumptions of Theorem \ref{prop1} correspond
to a row vector previously  
centered by $\nu$ and $\mu$, with signal $[ \psi_{1} \mathbf{1}_{(n_{1})}
\hspace{2mm} \psi_{2} \mathbf{1}_{(n_{2})} ]$, column covariance $\Delt$,
 and row variance $\sigma^{2}$, the diagonal element of $\Sig$.  For
 microarrays, the
 result states that 
when the columns (arrays) are correlated, the variance of the $Z$-statistic is
inflated or deflated by $\eta$, a function of the column covariance.
Notice that if $\Delt = \mathbf{I}$, $\eta = c_{n}$ and the variance
of $Z$ is one.  If there is only column correlation
within the two classes we have the following result.
\begin{corollary}
\label{cor_1}
Assume $\x = [ \x_{1} \hspace{2mm} \x_{2} ]$ with $\x_{1} \sim
N_{1,n_{1}}( 0, \psi_{1} \mathbf{1}_{(n_{1})}, \sigma^{2}, \Delt_{1} )$ and $\x_{2} \sim
N_{1,n_{2}} ( 0, \psi_{2} \mathbf{1}_{(n_{2})}, \sigma^{2}, \Delt_{2})$ such that
$\mathrm{Cov}(\x) = \Delt = \left( \begin{array}{cc} \Delt_{1} &
    \mathbf{0} \\ \mathbf{0} & \Delt_{2} \end{array} \right)$, then
\begin{align}
Z \sim N \left( \frac{ \psi_{1} - \psi_{2} }{ \sigma
  \sqrt{c_{n} }}, \frac{ \eta_{1} + \eta_{2} }{ c_{n}} \right)
\end{align}
$\textrm{  where } \eta_{k} \triangleq \displaystyle\frac{1}{n_{k}^{2}}
\displaystyle\sum_{i=1}^{n_{k}} \left( \sum_{j=1}^{n_{k}} L_{k,ij} \right)^{2}
= \frac{1}{n_{k}^{2}} \sum_{i=1}^{n_{k}} \sum_{j=1}^{n_{k}}
\Delt_{k,ij}$ for $k = 1, 2$ and $\mathbf{L}_{k}$ is the matrix
square root of $\Delt_{k}$.
\end{corollary}

In both of the previous results, we assumed that the row variance,
$\sigma^{2}$ was known.  However, in most microarray experiments this
is not known and must be estimated.  With $\sigma^{2}$ unknown, for
testing the hypothesis \eqref{hypoth}, the two-sample  
$t$-statistic is used.  
\begin{align}
\label{tstat}
T = \frac{\bar{x}_{1} - \bar{x}_{2}}{s_{\x_{1},\x_{2}}
  \sqrt{c_{n}}}, \hspace{6mm}  s^{2}_{\x_{1},\x_{2}} = \frac{
  \sum_{i \in y\mathcal{C}_{1}} \left( x_{1,i} - \bar{x}_{1}
  \right)^{2}  + \sum_{i \in \mathcal{C}_{2}} \left( x_{2,i} - \bar{x}_{2}
  \right)^{2} }{n_{1} + n_{2} - 2}
\end{align}
with $c_{n}$ and $\bar{x}_{k}$ as previously defined.
Under the null hypothesis, $T \sim t_{(n - 2)}$, while under the
alternative, $T \sim t( 
\delta)_{(n - 2)}$, a 
non-central $t$ distribution with non-centrality parameter $\delta =
( \psi_{1} - \psi_{2}) / (\sigma \sqrt{c_{n}})$.  

When there are column correlations as in the assumptions of Proposition
\ref{prop1}, however, the distribution of $T$ does not have a closed form. 
(The square of the pooled sample standard deviation is no longer distributed as a Chi-squared random
variable and the numerator and denominator of $T$ are not independent.)
Hence, we explore the effects of column correlations on the
$T$-statistic through a small simulation study.  Data is simulated
according to the assumptions of Theorem \ref{prop1} with $n = 50$
columns with $n_{1} = n_{2} = 25$ in each class.  Four structured
covariance matrices were used to assess the $Z$ and $T$-statistics
under the column correlations scenarios, as given below. 
{\footnotesize
\begin{itemize}
\item $\Delt_{1} : \Delt_{1,ij} =  0.9^{|i-j|}$.
\item $\Delt_{2}$: Blocked diagonal with blocks of size 10.
  Within each block, $\Delt_{2,ij} = 0.9^{|i-j|}$.
\item $\Delt_{3} : \Delt_{3,ij} =  0.5^{|i-j|}$.
\item $\Delt_{4}$: Blocked diagonal  with blocks of size 10.
  Within each block, $\Delt_{4,ij} = 0.5^{|i-j|}$.
\end{itemize}}

\begin{figure}[!!t]
\begin{center}
\includegraphics[width=5.5in]{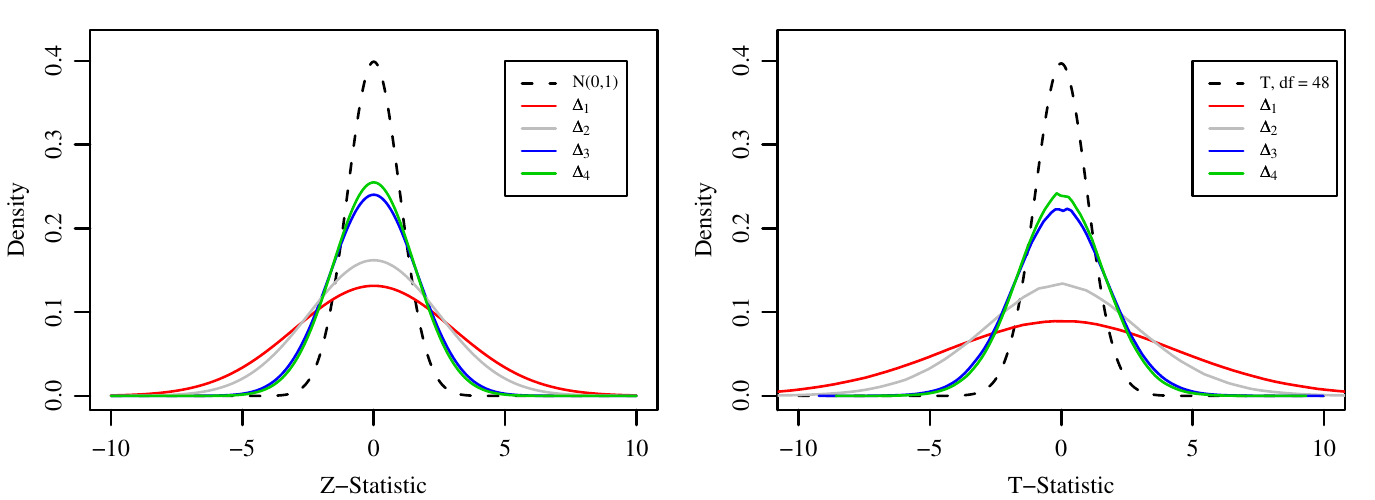}
\end{center}
\caption{\em \footnotesize Comparison of theoretical null distributions for the two-sample $Z$-statistic
  (left) and $T$-statistic (right) under various column covariance scenarios
  given in Section \ref{section_nulls}.  Variances of the
  $Z$-statistics were calculated by the result in Theorem
  \ref{prop1}, while the densities of the $T$-statistics were
  estimated via a simulation with one million replicates. 
\label{fig_array_cor}}
\end{figure}

Figure \ref{fig_array_cor} reveals the effect of column correlations on
the distributions of $Z$ and $T$.  We see that column correlations can
cause dramatic over-dispersion of the test statistics compared to their
theoretical null distribution.  This is a possible explanation to the
over-dispersion seen in the 
real microarray examples of Figure \ref{fig_tstat_micro}. Compared to
the variance of the $Z$-statistic, the $T$-statistic appears to be
even more affected by column correlations.  This is confirmed in Table
\ref{tab_null_var} where we present the variances of the
$Z$-statistic calculated by Theorem \ref{prop1} and the variances
of the $T$-statistic estimated by Monte Carlo simulation. Indeed,
small amounts of correlation in the columns 
can cause a dramatic increase in the variance of the $T$-statistic.

\begin{table}[!!t]
\begin{center}
\begin{tabular}{c|ll}
\hline
& Var($Z$-statistic) & Var($T$-statistic) \\ 
\hline
$\Delt_{1}$ & \hspace{3mm} 9.215 & 19.94 (0.029) \\
$\Delt_{2}$ & \hspace{3mm} 6.069 & \hspace{.5mm} 9.492 (0.0144) \\
$\Delt_{3}$ & \hspace{3mm} 2.76 & \hspace{.5mm} 3.197 (0.00472) \\
$\Delt_{4}$ & \hspace{3mm} 2.45 & \hspace{.5mm} 2.79 (0.00411) \\
\hline
\end{tabular} 
\end{center}
\caption{ \em \footnotesize Variances of the two-sample $Z$ and
  $T$-statistics under 
  various column correlation scenarios  as given in Section
  \ref{section_nulls}. Variances of the $T$-statistics were estimated
  via simulation with one million replicates. The theoretical variance
  of the $Z$-statistic should be one, and 1.022  for the $T$-statistic.
  \label{tab_null_var}}  
\end{table}

In this section, we have shown how the distribution of $T$ and
$Z$-statistics behave when columns or arrays are correlated.  When analyzing
microarrays, however, many have advocated using a non-parametric
method, estimating the null distribution by permutations \cite{dudoit_2003,
  storey_2003, tusher_2001}.  For the two-class microarray, one would
permute the class labels and calculate the $T$-statistic for each
permutation.  These 
permutations form a null distribution, as under the null hypothesis
\eqref{hypoth}, the class means are the same.  Thus, each permutation
of the labels is equally likely.  When the arrays are correlated,
however, this assumptions fails.  Each permutation of the columns is
not equally likely under the null due to the array covariance
structure.  While we do not explore the behavior of the permutation
nulls further in this section, we include
permutation-based methods from \citet{storey_2003} in our
simulation study in the following section.

\section{Study: Dependence and Multiple Testing}
\label{section_multiple}

In the previous section, we presented the theoretical null
distributions of commonly used test statistics for a single two-sample
test statistic when the columns are correlated.  With 
transposable data, however, one needs to test possibly tens of thousands of
row variables, thus creating a problem of multiplicity.   In this section, we
first review some multiple testing procedures that are known to
control errors under certain types of dependencies.  We then present a
series of simulations to study the behavior of commonly used multiple
testing procedures when the rows and columns are correlated.

\subsection{Background}

A common error measure for 
controling the number of false positives in microarrays is  the
False Discovery Rate (FDR).  This is the expectation of the False
Discovery Proportion (FDP): let $V$ be the number of false
positives and $R$ be the total number of rejections, then $q = FDR =
\mathrm{E} ( V/R | R > 0 )$.  Typically,
investigators seek to control the FDR at $q = 0.1$, meaning that on
average 10\% of rejections are false.  

The step-up method of \citet{bh_1995} is one of the most widely used
methods for controling the FDR.  \citet{benjamini_2001} have shown
that this method 
controls the FDR under types of positive dependence, specifically {\em
  positive regression dependence}, and  \citet{sarkar_2008} has relaxed
this assumption to slightly broader forms of positive
dependence.   This may 
not be appropriate for all types of transposable data, especially
microarrays where we expect some negative 
correlations between genes.  Alternatively, \citet{benjamini_2001}
have shown that dividing the
thresholds in the step-up procedure by a constant controls the FDR
under arbitrary dependencies.

Another commonly used method to control the FDR is
based on re-sampling or permutation distributions \citep{dudoit_2003,
  storey_2002, tusher_2001, yeuk_1999}.  Theoretically, these methods
are only known to 
control the FDR asymptotically under types of {\em weak dependence},
which encompasses forms of local dependence such as finite blocks
\citep{storey_2004}.  Thus, there could be many transposable data sets
in which the row variables do not satisfy these dependence structures.
(We also note that applying the step-up method to the
permutation-adjusted $p$-values is equivalent to the
direct FDR estimation  via re-sampling \citep{storey_2004}).

Also, to directly account for correlations, 
\citet{efron_2004, efron_2007} proposed a method to fit an
{\em empirical null} 
to the data.  One can then estimate the local FDR and then the FDR by
averaging the 
local FDR over the tail regions.

\subsection{Simulation Study}
\label{section_sim_study}

We study the effects of both row and column correlations on standard
statistical methodology used for 
large-scale inference through
a simulation study based on our matrix decomposition model.  We
compare FDR estimates of four types of 
FDR-controlling procedures to the true false
discovery proportion (FDP).  The four methods we compare are the
step-up method of \citet{bh_1995}, the step-up method for control
under arbitrary dependence of \citet{benjamini_2001}, the
permutation-based method of \citet{storey_2003}, and the
method based on the empirical null and local FDR's of
\citet{efron_2007}.  The two-sample $t$-statistic was used for
all methods with $p$-values computed by comparing it to the $t_{(n-2)}$
  distribution for the step-up procedures.   We used 1000 permutations for the 
permutation-based method.   The defaults in
the \textsf{localfdr} package available on \textsf{CRAN}, the
\textsf{R} language repository.  These defaults
fit the null distribution as a natural spline with seven degrees of
freedom, for the empirical null-based method.

Our simulation study is structured as follows.  The data is simulated
under the matrix decomposition model \eqref{decomp} and is of size
250 by 50.  The first 50 rows are non-null with a two-class signal
matrix given by $\psi_{1, 1:25} =  0.5$, $\psi_{1,26:50} = -0.5$,
$\psi_{2,1:25} = -0.5$, $\psi_{2,26:50} = 0.5$ and the last 200
elements of $\psi_{1}$ and $\psi_{2}$ equal to zero. 
We consider two types of row covariances, $\Sig_{1}$ with all positive
correlations satisfying the positive regression
dependence assumption of \citep{benjamini_2001}, and
$\Sig_{2}$ with both positive and negative correlations.  Both of these
row covariances are block diagonal.  We
simulate data under three column covariances, with the first being the
identity, or no column correlations.  The others, $\Delt_{1}$ and
$\Delt_{2}$ reflect a local and a class effect, respectively.
These simulation covariances are summarized below. 
{\footnotesize
\begin{itemize}
\item $\Sig_{1}$: Blocked diagonal with blocks of size 10.  Within
  each block, $\Sig_{1,ij} = 0.9^{|i-j|}$.
\item $\Sig_{2}$:  Blocked diagonal with blocks of size 10.  Within
  each block, $\Sig_{2,ij} = (-0.9)^{|i-j|}$.
\item $\Delt_{1}$:  Blocked diagonal with blocks of size 10.  Within
  each block, $\Delt_{1,ij} = 0.5^{|i-j|}$.
\item $\Delt_{2}$:  Blocked diagonal with blocks of size 25.  Within
  each block, $\Delt_{2,ij} = 0.5^{|i-j|}$. 
\end{itemize}}

\begin{figure}[!!t]
\begin{center}
  \includegraphics[width=5in]{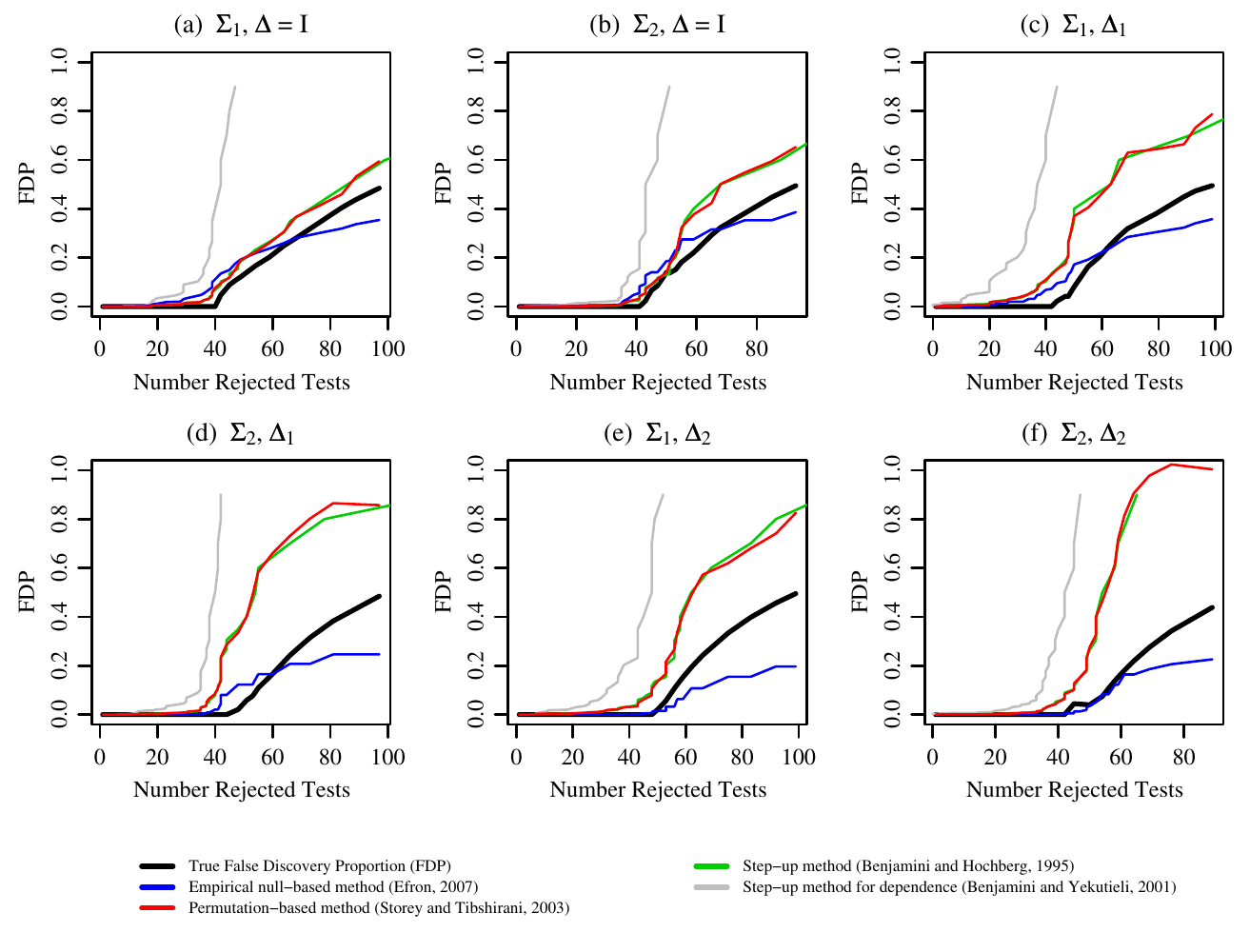}
\end{center}
\caption{\em  \footnotesize Simulation Study FDR Curves: The true and
  estimated false 
discovery proportions plotted against the number of tests rejected for
each of the six simulations.  All data was simulated under the
matrix decomposition model, \eqref{decomp}, with parameters given
in Section \ref{section_sim_study}. 
\label{fig_sim_study}}
\end{figure}

We present plots of the true FDP verses the
number of hypotheses rejected for the four methods for one realization
of each of the six
simulation scenarios in Figure \ref{fig_sim_study}.  We note that the
lines above the true FDP curve denote conservative FDR estimates.  
In Table \ref{tab_sim_study}, we report the true and estimated false
discovery proportions (FDP) when fixed numbers of hypotheses are
rejected (40, 45, 50, 55 and 60 tests).  Results are averaged over ten
simulations with the standard error also reported.

\begin{table}[!!t]
\begin{center}
\scalebox{.8}{
\begin{tabular}{r|l|llll}
\hline
& \multirow{2}{*}{True FDP} & (Benjamini \& & (Benjamini \& &
(Storey \& & \multirow{2}{*}{(Efron, 2007)} \\
& &  Hochberg, 1995) &  Yekutieli, 2001) &
 Tibshirani, 2003) &  \\
\hline
(a) $\Sig_{1}, \Delt = \mathbf{I}$ \hspace{6mm}  &&&&& \\
\hspace{6mm} 40 tests & 0.0725 (0.033) & 0.0723 (0.022) & 0.387 (0.091) & 0.0742 (0.022) &
0.257 (0.068) \\ 
\hspace{6mm} 45 tests & 0.104 (0.034) & 0.103 (0.026) & 0.545 (0.11) & 0.105 (0.026) &
0.286 (0.07) \\ 
\hspace{6mm} 50 tests & 0.124 (0.033) & 0.147 (0.028) & 0.715 (0.1) & 0.149 (0.029) & 0.32
(0.069) \\ 
\hspace{6mm} 55 tests & 0.169 (0.025) & 0.19 (0.03) & 0.823 (0.092) & 0.193 (0.03) & 0.344
(0.066) \\ 
\hspace{6mm} 60 tests & 0.222 (0.02) & 0.255 (0.036) & 0.891 (0.069) & 0.258 (0.036) &
0.376 (0.063)\\ 
\hline
(b) $\Sig_{2}, \Delt = \mathbf{I}$ \hspace{6mm}  &&&&& \\
\hspace{6mm} 40 tests & 0.035 (0.017) & 0.0498 (0.0081) & 0.304 (0.049) & 0.0513 (0.0082) & 0.161 (0.034) \\
\hspace{6mm} 45 tests & 0.0711 (0.019) & 0.0839 (0.012) & 0.512 (0.076) & 0.0856 (0.012) & 0.209 (0.041) \\
\hspace{6mm} 50 tests & 0.12 (0.018) & 0.141 (0.021) & 0.771 (0.099) & 0.143 (0.021) & 0.249 (0.045) \\
\hspace{6mm} 55 tests & 0.167 (0.016) & 0.191 (0.029) & 0.822 (0.094) & 0.192 (0.029) & 0.278 (0.04) \\
\hspace{6mm} 60 tests & 0.217 (0.014) & 0.243 (0.035) & 0.867 (0.074) & 0.246 (0.035) & 0.311 (0.041) \\
\hline
(c) $\Sig_{1}, \Delt_{1}$ \hspace{10mm}  &&&&& \\
\hspace{6mm} 40 tests & 0.0075 (0.0053) & 0.0588 (0.019) & 0.329 (0.087) & 0.0578 (0.019) & 0.0152 (0.0053) \\
\hspace{6mm} 45 tests & 0.0222 (0.0099) & 0.134 (0.037) & 0.599 (0.11) & 0.133 (0.038) & 0.0511 (0.025) \\
\hspace{6mm} 50 tests & 0.056 (0.016) & 0.245 (0.052) & 0.847 (0.087) & 0.247 (0.053) & 0.0858 (0.032) \\
\hspace{6mm} 55 tests & 0.111 (0.011) & 0.426 (0.04) & 1 (0) & 0.43 (0.041) & 0.139 (0.032) \\
\hspace{6mm} 60 tests & 0.178 (0.0083) & 0.579 (0.052) & 1 (0) & 0.585 (0.053) & 0.167 (0.028) \\
\hline
(d) $\Sig_{2}, \Delt_{1}$ \hspace{10mm}  &&&&& \\
\hspace{6mm} 40 tests & 0.005 (0.005) & 0.0455 (0.0065) & 0.277 (0.04) & 0.0439 (0.0065) & 0.00846 (0.0041) \\
\hspace{6mm} 45 tests & 0.0111 (0.005) & 0.111 (0.027) & 0.58 (0.09) & 0.11 (0.026) & 0.0198 (0.0076) \\
\hspace{6mm} 50 tests & 0.042 (0.0081) & 0.225 (0.046) & 0.869 (0.082) & 0.225 (0.047) & 0.0493 (0.014) \\
\hspace{6mm} 55 tests & 0.109 (0.0086) & 0.404 (0.034) & 1 (0) & 0.409 (0.034) & 0.0923 (0.017) \\
\hspace{6mm} 60 tests & 0.178 (0.0056) & 0.552 (0.048) & 1 (0) & 0.554 (0.048) & 0.133 (0.018) \\
\hline
(e) $\Sig_{1}, \Delt_{2}$ \hspace{10mm}  &&&&& \\
\hspace{6mm} 40 tests & 0.0125 (0.0077) & 0.0831 (0.018) & 0.476 (0.09) & 0.0783 (0.019) & 0.0749 (0.024) \\
\hspace{6mm} 45 tests & 0.0333 (0.015) & 0.164 (0.031) & 0.746 (0.097) & 0.16 (0.032) & 0.117 (0.032) \\
\hspace{6mm} 50 tests & 0.078 (0.015) & 0.281 (0.027) & 0.969 (0.031) & 0.276 (0.028) & 0.165 (0.032) \\
\hspace{6mm} 55 tests & 0.135 (0.012) & 0.368 (0.028) & 1 (0) & 0.364 (0.028) & 0.194 (0.028) \\
\hspace{6mm} 60 tests & 0.198 (0.011) & 0.461 (0.044) & 1 (0) & 0.458 (0.045) & 0.234 (0.028) \\
\hline
(f) $\Sig_{2}, \Delt_{2}$ \hspace{10mm}  &&&&& \\
\hspace{6mm} 40 tests & 0.0075 (0.0053) & 0.0712 (0.016) & 0.407 (0.073) & 0.066 (0.015) & 0.0444 (0.012) \\
\hspace{6mm} 45 tests & 0.0311 (0.011) & 0.169 (0.022) & 0.855 (0.072) & 0.163 (0.021) & 0.087 (0.019) \\
\hspace{6mm} 50 tests & 0.078 (0.012) & 0.277 (0.025) & 0.99 (0.0095) & 0.271 (0.025) & 0.132 (0.021) \\
\hspace{6mm} 55 tests & 0.144 (0.013) & 0.388 (0.037) & 1 (0) & 0.383 (0.037) & 0.167 (0.019) \\
\hspace{6mm} 60 tests & 0.198 (0.013) & 0.47 (0.044) & 1 (0) & 0.468 (0.043) & 0.197 (0.02) \\
\hline
\end{tabular} }
\end{center}
\caption{\em  \footnotesize  Simulation Study: The effect of row and column correlations
  on estimation of the false discovery rates.  The true false
  discovery proportion (FDP) and estimates with standard errors using
  the step-up, step-up 
  for dependence, permutation, and empirical null based methods, as
  described in Section \ref{section_sim_study}, are given when a
  pre-specified number of tests are rejected.  All simulations were done
using the matrix decomposition model, \eqref{decomp}, with parameters
given in Section \ref{section_sim_study}, and repeated ten times. 
\label{tab_sim_study}}
\end{table}

This simulation study reveals several interesting results. 
First, dependencies among the rows do not seem to effect FDR
estimation with the four
multiple testing procedures.  When, $\Delt = \mathbf{I}$ as in
simulations (a) and (b),  the methods
generally conservatively estimate the true FDP.  This is noteworthy
since besides the method of \citet{benjamini_2001}, there are limited
theoretical results supporting FDR control under various dependencies.

When there are even moderate correlations between columns, simulations
(c) through (f), the
four methods give poor estimates of the FDR.  The step-up method and
the permutation-based method perform similarly.  They both give
extremely conservative estimates of the FDP when there is either a
local or a class effect among the columns.  Thus, when using
these methods for controlling the FDR at $q = 0.1$, for example, one
would reject less than 45 genes, when in reality one should be
permitted to reject around 55 genes.  We also see that while the
method of \citet{benjamini_2001} controls the FDR are arbitrary
dependencies, in practice this method is much too conservative for
general use.  On the other hand, the empirical null-based method of
\citet{efron_2007} performs inconsistently.

Overall, the results of this simulation study reveal that dependencies
among rows do not seem to effect the performance of the multiple
testing procedures.  On the other hand, the theoretical results of
Section \ref{section_nulls} are confirmed: dependencies among the columns
are extremely problematic when conducting large-scale inference.

\section{De-Correlating a Matrix}
\label{section_de_cor}

In the previous sections, we have presented theoretical and simulation
results demonstrating some of the problems with using standard
statistical methodology for making inferences on transposable data.
In the remainder of this 
paper, we present a solution to these problems by directly estimating the
covariances and using these to {\em sphere} or de-correlate
the data.  

The key to our method, based on the matrix
decomposition model \eqref{decomp}, is the simultaneous estimation of
the row and column covariances.
 This is important because of the
close relationship between the observed row and column covariances.
Take, for example, the empirical covariances of a 
centered data matrix $\X$, $\hat{\Sig} = \X \X^{T}/m$ and $\hat{\Delt}
= \X^{T} \X / n$.  If we take the singular value decomposition, $\X =
\U \D \V^{T}$, then $\hat{\Sig} = \U \D \U^{T}$ and $\hat{\Delt} = \V
\D \V^{T}$, i.e. the two covariances estimates share the same
eigenvalues.  In fact, \citet{efron_2009_indep} shows that the variance of
the two correlation matrices is the same.  Because of this, population
correlations among the 
rows, for example, often make the columns seem correlated.  Thus,
estimating the column covariance without accounting for the
covariances of the rows is problematic.  With Transposable
Regularized Covariance Models, we can estimate both $\Sig$ and $\Delt$
simultaneously according the the matrix-variate normal framework.  We
review these models and discuss their relevance for the example of
microarrays in the next section.

\subsection{Review: Transposable Regularized Covariance Models}

The Transposable Regularized Covariance Model (TRCM) allows us to
estimate a non-singular row and column covariance matrix by maximizing
a penalized log-likelihood of the matrix-variate normal distribution
\citep{allen_2010}.  The model places a strictly convex penalty on the
inverse covariances, or concentration matrices, of the rows and
columns.  For estimating the covariances in this context, we propose to
use a sparsity-inducing penalty, an $L_{1}$ penalty, on the
concentration matrices.  Following from the matrix decomposition
model, \eqref{decomp}, if we let $\N$ be the 
noise matrix remaining after removing the means and the signal in the
data, then the penalized log-likelihood is as follows.  
\begin{align}
\label{trcm}
\ell(\Sig, \Delt) = \frac{n}{2} \mathrm{log} | \Sigi | + \frac{m}{2}
\mathrm{log} | \Delti | - \frac{1}{2} \mathrm{tr} \left( \Sigi \N \Delti \N^{T}
\right) - \lambda m || \Sigi ||_{1} - \lambda n || \Delti ||_{1} 
\end{align}
where $|| \Delti ||_{1} = \sum_{i=1}^{n} \sum_{j=1}^{n} | \Delti_{ij}
|$ and $\lambda$ is a penalty parameter that must be estimated.

We motivate the use of \eqref{trcm} first by discussing practical
considerations.  As the columns of the data matrix are usually
There are several advantages assumed to be independent, $\Delt =
\mathbf{I}$, this should be our default position.    By placing an
$L_{1}$ penalty on $\Delti$, our 
model encourages sparsity in the off-diagonal elements of
$\Delt$.  Also, notice that we have one penalty parameter, $\lambda$,
that is modulated by the dimension of the rows and columns. (We note
that the penalty parameter, $\lambda$, can be 
selected by cross-validation). 
Thus, the evidence of a partial correlation among columns must be
strong relative to the correlations among the rows for an
off-diagonal element of $\Delti$ to be estimated as non-zero.
Secondly,specifically for microarrays, it seems reasonable to assume
that the covariances among the 
genes is sparse, since biologically genes are likely only to be
correlated with genes in the same or related pathways.

We also pause briefly to discuss the theoretical rationale for using
$L_{1}$ penalties, instead of, for example, $L_{2}$ penalties.
Recall that covariance solutions to the TRCM model with $L_{2}$
penalties have eigenvectors that are equal to the left and right
singular vectors of the data \citep{allen_2010}.  Thus, the singular
vectors of the data would remain the same when sphering with these
estimates.  In high-dimensional
settings, however, it is well established that eigenvectors of 
empirical covariances are inconsistent \citep{johnstone_2004}, and
thus, sphering by the $L_{2}$ covariance estimates seems ill-advised.
While the consistency of $L_{1}$ TRCM estimates has not been
established, there are consistency results for multivariate covariance
estimation with an $L_{1}$ penalty.  \citet{rothman_2008} show convergence of
the multivariate covariance estimate in the Frobenius norm and more
importantly for the correlation estimate in the operator norm which
implies convergence of the eigenvectors \citep{el_karoui_2008}.  These
results reveal some of the possible theoretical advantages of using
$L_{1}$ penalties to estimate the covariances.

\subsection{Sphering Algorithm}
\label{section_sphering}

Based on the matrix decomposition model, \eqref{decomp}, we
present a method of de-correlating or sphering the data so that
the rows and columns are approximately independent.  This sphered
data can then be used with standard multiple testing procedures
to identify significant row variables.  Given a
data matrix $\X$ with $m$ rows and $n$ columns, we
present our sphering algorithm in Algorithm \ref{sphering_alg}.
{\small
\begin{algorithm}
\caption{Sphering Algorithm}
\label{sphering_alg}
\begin{enumerate}
\item Estimate row and column means, $\hat{\nu}$ and $\hat{\mu}$ forming
  $\hat{\M}$, and the signal matrix, $\hat{\Smat}$.
\item Define the noise, $\N \triangleq \X - \hat{\M} - \hat{\Smat}$.  Estimate row and column covariances of noise, $\hat{\Sig}$ and
  $\hat{\Delt}$ via TRCM.
\item Sphere the noise: $\tilde{\N} \triangleq \hat{\Sig}^{-\frac{1}{2}} \N
  \hat{\Delt}^{-\frac{1}{2}}$.   Form the sphered data matrix: $\tilde{\X}
  \triangleq \hat{\Smat} + \tilde{\N}$. 
\end{enumerate}
\end{algorithm}}

The Sphering Algorithm simply estimates the means and signal
according to the matrix decomposition model \eqref{decomp} and
then estimates the correlation structure among the rows and columns in
the remaining noise.  The TRCM covariance estimates, $\hat{\Sig}$ and
$\hat{\Delt}$ are used to de-correlated the noise.  Here,
$\hat{\Sig}^{-1/2}$ is the matrix square root of $\hat{\Sig}^{-1}$
and $\hat{\Delt}^{-1/2}$ of
$\hat{\Delt}$.  (We use the symmetric square root defined by the
following.  Let $\hat{\Sig}^{-1} = \mathbf{P} \Lambda \mathbf{P}^{T}$ be
the eigenvalue decomposition of $\hat{\Sig}^{-1}$, then the symmetric
matrix square root is given by $\hat{\Sig}^{-1/2} = \mathbf{P}
\Lambda^{1/2} \mathbf{P}^{T}$.)  Adding the
signal back into this sphered noise, we obtain $\tilde{\X}$ which we call
the sphered data.  One can use this de-correlated data to
find significant row variables. 

Now, we investigate some of the theoretical properties of the sphered
data for the two-class problem introduced in Section
\ref{section_nulls}.  The sphered data, $\tilde{\X}$ has the
following properties.
\begin{proposition}
\label{sphered_dist}
Let $\X \sim   N_{m,n} \left( \M + \Smat, \Sig , \Delt  \right)$
where $\M =  \nu \mathbf{1}_{(n)}^{T} + \mathbf{1}_{(m)} \mu^{T} $ and
$\Smat = [ \psi_{1} \mathbf{1}_{(n_{1})}^{T} \hspace{2mm}  \psi_{2}
\mathbf{1}_{(n_{2})}^{T}  ] $  and let $\tilde{\X}$ be the sphered
data given by Algorithm 
\ref{sphering_alg}.   Then, 
\begin{itemize}
\item[(i)] $\mathrm{E}(\tilde{\X}) = \Smat = [ \psi_{1} \mathbf{1}_{(n_{1})}
  \hspace{2mm} \psi_{2} \mathbf{1}_{(n_{2})} ]$,
\item[(ii)] $\tilde{\X}  - \hat{\Smat} \sim N_{m,n} \left( \mathbf{0},
    \mathbf{0}, \tilde{\Sig}, \tilde{\Delt} \right)$, 
\end{itemize}
where $\tilde{\Sig} = \hat{\Sig}^{-\frac{1}{2}} \Sig
\hat{\Sig}^{-\frac{1}{2}}$ and $\tilde{\Delt} = \hat{\Delt}^{-\frac{1}{2}} \Delt
\hat{\Delt}^{-\frac{1}{2}}$.
\end{proposition}

Thus, the class signal remains the same between $\X$ and $\tilde{\X}$,
and the covariance structure is all that changes.  By sphering the
noise, the noise of each row in $\tilde{\X}$ becomes a linear
combination of the noise in the other rows.  

We now study how sphering the data affects the $Z$ and $T$
statistics from Section \ref{section_nulls}.  First, the $Z$-statistic
does not change with sphering.  The numerator of
both the $Z$ and $T$ statistic, $\bar{x}_{1} - \bar{x}_{2}$ is given
by $\hat{\psi}_{1} - \hat{\psi}_{2}$, the components of the estimated
signal matrix $\hat{\Smat}$.  The
denominator of the $T$-statistic, namely $s_{x_{1},x_{2}}$, the
estimate of the noise, however, changes with sphering.  Recall that in
Section \ref{section_nulls}, we discussed how the $T$-statistic does
not have a closed form distribution when there are column correlations.
After sphering the data, however, the $T$-statistic on the sphered
data follows a scaled $t$ distribution under certain conditions.  This
is given by the following result.  
\begin{proposition}
\label{sphering_t}
Let $\X \sim   N_{m,n} \left( \M + \Smat, \Sig , \Delt  \right)$
where $\M =  \nu \mathbf{1}_{(n)}^{T} + \mathbf{1}_{(m)} \mu^{T} $ and
$\Smat = [\psi_{1} \mathbf{1}_{(n_{1})}^{T} \hspace{2mm}  \psi_{2}
\mathbf{1}_{(n_{2})}^{T}  ] $.  Let $\tilde{\X}$ be the sphered  
data given by Algorithm 
\ref{sphering_alg}, and let the statistic $\tilde{T}_{i}$ be the 
statistic for the $i^{th}$ row defined by \eqref{tstat} for the data
$\tilde{\X}$.  Then under the null hypothesis $H_{0}: \psi_{1} = \psi_{2}$, \\
\begin{center} if $\tilde{\Delt} = \mathbf{I}$,\hspace{4mm} $\tilde{T}_{i} \sim
   \displaystyle\frac{\tilde{\sigma_{i}}}{\sigma_{i}}
   \sqrt{\displaystyle\frac{\eta}{c_{n}}}   t_{(n-2)}$,\\
\end{center}  
  where $c_{n} = \frac{1}{n_{1}} + \frac{1}{n_{2}}$, $\eta =
  \sum_{j=1}^{n} \left( \frac{1}{n_{1}} \sum_{i \in 
    \mathcal{C}_{1}} L_{ij} - \frac{1}{n_{2}} \sum_{i \in
    \mathcal{C}_{2}} L_{ij}  \right)^{2}$ and 
 $\mathbf{L}$ is the matrix square root of $\Delt$, $\sigma_{i}
  = \Sig_{ii}$ and $\tilde{\sigma}_{i} = \tilde{\Sig}_{ii}$. 
\end{proposition}

Using our sphering algorithm, we obtain test statistics that
follow known distributions when the sphered column covariance,
$\tilde{\Delt}$ is the identity.  If $\tilde{\Delt}$ is instead
a diagonal matrix, then a simple scaling of the columns will give the
above result.  Notice that if the original data, $\X$,
has no column correlations, $\Delt = \mathbf{I}$, and
$\tilde{\sigma}_{i} = \sigma_{i}$, then $T$ and $\tilde{T}$ both follow a
$t$ distribution with $n-2$ degrees of freedom.  Thus, if the data
originally follows the correct theoretical null distribution, then sphering the
data does not change its null distribution.  Also, if the sphered rows
are independent, $\tilde{\Sig} = \mathbf{I}$, or approximately
independent, then the statistics, $T_{i}$ are independent or
approximately independent.  We also note that we can often assume that
$\tilde{\sigma}_{i} = \sigma_{i}$, thus eliminating that coefficient ratio from
the distribution.  This is especially a reasonable assumption if the
rows are scaled prior to applying the sphering algorithm.

Our results in Proposition
\ref{sphering_t} hold if the sphered column covariance $\tilde{\Delt}$
is the identity or diagonal.  The TRCM model, however, estimates sparse
penalized row and column covariances.  These penalized estimates will
not capture the full covariances, but will instead estimate the major
correlations.  Thus, in practice, $\tilde{\Delt}$ and $\tilde{\Sig}$
are not likely to be exactly the identity.  
We have observed in simulations, however, that $\tilde{\Delt}$ is
often diagonal or nearly diagonal, and thus the theoretical results
are appropriate.

When calculating $p$-values for $\tilde{T}$ based on the distribution
given in Proposition \ref{sphering_t}, we must know the value of
$\eta$ which depends on the original column covariance $\Delt$.  One
could estimate this from $\hat{\Delt}$, but since the TRCM
framework estimates penalized covariances, an estimate, 
$\hat{\eta}$, based on $\hat{\Delt}$ will underestimate the population
$\eta$.   Hence to obtain the null distribution of the
$\tilde{T}$-statistics,  we have 
opted to scale $\tilde{T}$ by the variance of the central portion
of the observed distribution of the test statistics.  This procedure is
outlined in Algorithm \ref{alg_cen_matching} where $\rho_{\alpha}(x)$
denotes the 
$\alpha^{th}$ quantile of $x$. and $I()$ is the indicator function.  
\begin{algorithm}
\label{alg_cen_matching}
\caption{Scaling by the central portion of $\tilde{T}$.}
\begin{enumerate}
\item Let the expected proportion of null test statistics be 
  $\hat{\pi}_{0} = \hat{m}_{0}/m$.
\item Estimate the variance of the central portion of sphered test
  statistics:
\vspace{-2mm}
\begin{align*}
\hat{\sigma}^{2}_{\tilde{T}}(\hat{\pi}_{0}) \triangleq \mathrm{\hat{Var}} \left[
  \tilde{T}_{i} \hspace{2mm} \mathrm{I} \left( \tilde{T}_{i} \geq \rho_{\left(
      (1-\hat{\pi}_{0})/2\right)}(\tilde{T}), 
\tilde{T}_{i}  \leq \rho_{\left( 1 - \hat{\pi}_{0}/2 \right)}(\tilde{T})
\right) \right]
\end{align*}
\vspace{-2mm}
\item Define the central-matched $\tilde{T}$-statistics:
  $\tilde{T}^{*} \triangleq \tilde{T}
  \sigma_{t_{(n-2)}}(\hat{\pi}_{0}) /
  \hat{\sigma}_{\tilde{T}}(\hat{\pi}_{0})$,   
 where  $\sigma^{2}_{t_{(n-2)}}(\hat{\pi}_{0})$ is the variance of the central
  portion of the $t_{(n-2)}$ distribution.
\end{enumerate}
\end{algorithm}

We scale by the central portion of the $\tilde{T}$-statistics so that
the statistics can be tested against the
$t_{(n-2)}$ distribution.  Notice that if all of the test statistics
are null and 
$\pi_{0} = 1$ then, central-matching the variances reduces to scaling the
$\tilde{T}$-statistics.  Since under the assumptions of Proposition
\ref{sphering_t}, only the null $\tilde{T}_{i}$ follow a scaled
$t$-distribution, we do not want statistics corresponding to non-null
tests to contaminate the variance estimates.  Thus, we recommend using
a conservative estimate of $\pi_{0}$, such as 0.8 or 0.9 for
microarrays.

By applying our sphering algorithm, we directly account for
correlations among the rows and 
columns.   This results in test statistics that more closely follow
both their theoretical nulls and the theoretical
assumptions under which common multiple testing procedures are known
to control the false discovery rate.

\section{Results}
\label{section_res}

We now evaluate the performance of our sphering algorithm through many
simulated examples.  First, we compare data pre-processed by sphering
to the standard row and column centering method on simulations based on the
matrix decomposition model, \eqref{decomp}.  We use simulations from
the matrix-variate 
normal with the structured covariances from the simulation study in
Section \ref{section_sim_study} and also with covariances based on the
observed dependencies in real microarray data.  Finally, we test the
robustness of our method and compare it to other methods for modeling
dependencies in Section \ref{section_sim_other}.  For
all simulations, the sphering algorithm was applied with the TRCM
penalty parameter $\lambda$ selected by five-fold cross-validation and
with statistics scaled by the central portion using $\pi_{0} =
0.8$. (Note that the ``standard'' pre-processing method refers to row
and column centering throughout this section.)

\subsection{Simulations: Matrix-variate Model}
\label{section_sim_mv}

In all of these simulations, the data, $\X$ is simulated
from the matrix decomposition model \eqref{decomp} with $m = 250$
rows and $n = 50$ columns.  The first 50
rows are non-null given by $\psi_{1, 1:25} =  0.5$, $\psi_{1,26:50} =
-0.5$, 
$\psi_{2,1:25} = -0.5$, $\psi_{2,26:50} = 0.5$ and the last 200
elements of $\psi_{1}$ and $\psi_{2}$ equal to zero.

\begin{table}[!!t]
\begin{center}
\scalebox{.8}{
\begin{tabular}{r|l|llll}
\hline
& & \multicolumn{3}{c}{FDR Estimates} \\
& \multirow{2}{*}{True FDP} & (Benjamini \& & (Storey \& &
\multirow{2}{*}{(Efron, 2007)} \\ 
& &  Hochberg, 1995) &   Tibshirani, 2003) &  \\
\hline
$\Sig_{1}, \Delt = \mathbf{I}$ \hspace{6mm}  &&&&& \\
\multirow{2}{*}{\hspace{6mm} 40 tests} & 0.0725 (0.033) & 0.0723 (0.022) &  0.0742 (0.022) & 0.257 (0.068) \\
 &  {\bf 0.0333 (0.017)} & {\bf 0.0458 (0.019)} &  {\bf 0.0452 (0.019)} & {\bf 0.153 (0.075)} \\
\cline{2-5}
\multirow{2}{*}{\hspace{6mm} 45 tests} & 0.104 (0.034) & 0.103 (0.026) &  0.105 (0.026) & 0.286 (0.07) \\
 & {\bf 0.0469 (0.02)} & {\bf 0.0703 (0.025)} &  {\bf 0.0705 (0.025)} & {\bf 0.173 (0.076)} \\
\cline{2-5}
\multirow{2}{*}{\hspace{6mm} 50 tests} & 0.124 (0.033) & 0.147 (0.028) &  0.149 (0.029) & 0.32 (0.069) \\
 & {\bf 0.0822 (0.02)} & {\bf 0.104 (0.029)} &  {\bf 0.105 (0.029)} & {\bf 0.207 (0.075)} \\
\cline{2-5}
\multirow{2}{*}{\hspace{6mm} 55 tests} & 0.169 (0.025) & 0.19 (0.03) &  0.193 (0.03) & 0.344 (0.066) \\
 & {\bf 0.141 (0.016)} & {\bf 0.185 (0.035)} &
 {\bf 0.186 (0.035)} &
 {\bf 0.261 (0.067)} \\
\cline{2-5}
\multirow{2}{*}{\hspace{6mm} 60 tests} & 0.222 (0.02) & 0.255 (0.036) &  0.258 (0.036) & 0.376 (0.063) \\
 & {\bf 0.194 (0.012)} & {\bf 0.233 (0.038)} &  {\bf 0.234 (0.038)} & {\bf 0.284 (0.067)} \\
\hline
$\Sig_{1}, \Delt_{1}$ \hspace{10mm}  &&&&& \\
\multirow{2}{*}{\hspace{6mm} 40 tests} & 0.0075 (0.0053) & 0.0588 (0.019) &  0.0578 (0.019) & 0.0152 (0.0053) \\
 & {\bf 0.00278 (0.0028)} & {\bf 0.00493 (0.0029)} &  {\bf 0.00469 (0.0029)} & {\bf 0.0071 (0.0049)} \\
\cline{2-5}
\multirow{2}{*}{\hspace{6mm} 45 tests} & 0.0222 (0.0099) & 0.134 (0.037) &  0.133 (0.038) & 0.0511 (0.025) \\
 & {\bf 0.00988 (0.0099)} & {\bf 0.0157 (0.0071)} &  {\bf 0.0152 (0.007)} & {\bf 0.0171 (0.0086)} \\
\cline{2-5}
\multirow{2}{*}{\hspace{6mm} 50 tests} & 0.056 (0.016) & 0.245 (0.052) &  0.247 (0.053) & 0.0858 (0.032) \\
 & {\bf 0.0222 (0.011)} & {\bf 0.0438 (0.013)} &  {\bf 0.0434 (0.013)} & {\bf 0.0487 (0.018)} \\
\cline{2-5}
\multirow{2}{*}{\hspace{6mm} 55 tests} & 0.111 (0.011) & 0.426 (0.04) &  0.43 (0.041) & 0.139 (0.032) \\
 & {\bf 0.105 (0.0085)} & {\bf 0.124 (0.02)} &  {\bf 0.125 (0.02)} & {\bf 0.118 (0.018)} \\
\cline{2-5}
\multirow{2}{*}{\hspace{6mm} 60 tests} & 0.178 (0.0083) & 0.579 (0.052) &  0.585 (0.053) & 0.167 (0.028) \\
 & {\bf 0.172 (0.0039)} & {\bf 0.199 (0.028)} &  {\bf 0.201 (0.029)} & {\bf 0.146 (0.016)} \\
\hline
\end{tabular}
} 
\end{center}
\caption{\em \footnotesize  A subset of the simulation study results: True false
  discovery proportions (FDP) and FDR estimates with standard 
  errors are given when a 
  pre-specified number of tests are rejected.    Results using the
  sphering algorithm (in bold) are compared to data
  that has been row and column centered.  All data was simulated 
under the matrix decomposition model, \eqref{decomp}, with parameters
given in Section \ref{section_sim_study}, and repeated ten times.  Two
sets of values should be compared: the true FDP with sheering to
without sphering, and the FDR estimates compared to the true FDP for
both with and without sphering.
\label{sim_study_res}}
\end{table}

In Table
\ref{sim_study_res}, we present results on a subset of the simulations
from our simulation study in Section \ref{section_sim_study}.  The
remaining simulation study results are given in Appendix
\ref{section_app_sims}.  
The results in Table \ref{sim_study_res} show
that de-correlating the data matrix yields improvements 
in 1) statistical power and 2) estimation of the FDR.  We
briefly illustrate this by examining a specific example from Table
\ref{section_sim_study}.  Take the simulation with parameters
$\Sig_{1}$, $\Delt_{1}$ and look at the results with 55 tests
rejected.  We notice that the true FDP for the data pre-processed by
the standard method is 
0.111 whereas it is lower, 0.105, on the data that was sphered.  This
results from a favorable re-ordering of the test statistics that
gives a higher statistical power, one minus the true FDP.  Next,
notice that the FDR estimates for the step-up and permutation-based
methods are 0.426 and 0.43 respectively for the un-sphered data.
These estimates are overly conservative, as the true FDP is 0.111.
After sphering, however, the FDR estimates are 0.124 and 0.125 which
are much closer to the true FDP of 0.105.  Hence, sphering also
improves FDR estimation.

\begin{figure}[!!t]
\begin{center}
  \includegraphics[width=5.5in]{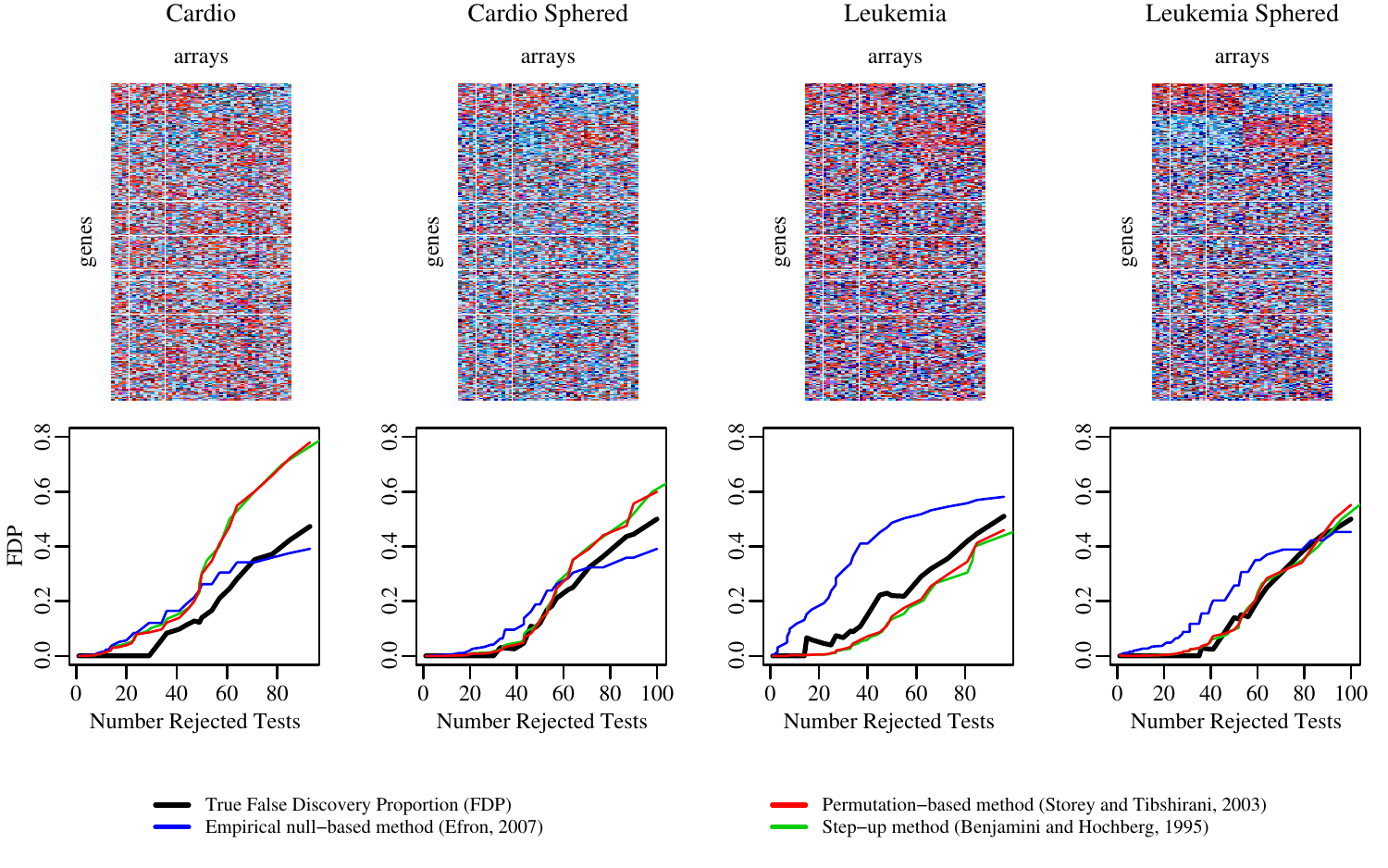}
\end{center}
\caption{ \em \footnotesize Example data images (top panel) and FDR curves (bottom panel)
  for the simulations based on dependencies within the ``Cardio'' and
  ``Leukemia'' microarrays.  Data is either gene and array centered or
  sphered.   In the FDR curves, the true and
  estimated false 
  discovery proportions are plotted against the number of genes
  rejected.  All data was simulated under the 
  matrix decomposition model, \eqref{decomp}, with parameters given
  in Section \ref{section_sim_mv}.
\label{fig_res_micro}}
\end{figure}

Sphering the data as a pre-processing step to multiple testing
procedures has many advantages.  First in microarrays, the higher
statistical power 
that results from a re-ordering of test statistics is important to
scientists who desire 
the top ranked genes from one microarray study to translate to the top
genes in another study.  Also, while sphering leads to improvements in
FDR estimation, it 
is still a slightly conservative estimate, as desired, for the true
FDP.  As with 
the simulation study, we find that the empirical null based method of
\citet{efron_2007} gives an inconsistent estimate of the FDR as it is
both a conservative and liberal estimate for differing numbers of
rejected tests.

We also wish, however, to test the performance
of our sphering algorithm on data with dependencies more similar to
real microarray data.  Thus, we build a second simulation
study based upon the empirical covariances of the ``Cardio'' and the
``Leukemia'' microarrays.  For each of the 
ten repetitions, we sample 250 genes and 50 arrays at random from each
microarray.  Let $i \in \mathcal{I}$ and $j \in
\mathcal{J}$, be the sampled sets of the genes and arrays
respectively, and assume $\X$ is the centered data matrix.  We then
calculate the empirical  
covariances, $\Delt_{MLE} = \sum_{i \in \mathcal{I}} \sum_{i' \in
  \mathcal{I}} X_{i}^{T} X_{i'} / m$, and  $\Sig_{MLE} = \sum_{j \in
  \mathcal{J}} \sum_{j' \in  \mathcal{J}} X_{j} X_{j'}^{T} / n$.  The
 data is simulated from the mean-restricted matrix-variate normal with
 $\X \sim N( \left[ \psi_{1} \hspace{2mm} \psi_{2} \right], \mathbf{0},
 \Sig_{MLE}, \Delt_{MLE})$.  Hence, the
 simulated data follows the observed covariance of the ``Cardio'' and
 ``Leukemia'' studies.   Example images and FDR curves from this
 simulation are given in Figure \ref{fig_res_micro} as well as the
 simulation results in Table \ref{tab_res_micro}.

\begin{table}[!!t]
\begin{center}
\scalebox{.8}{
\begin{tabular}{r|l|llll}
\hline
& & \multicolumn{3}{c}{FDR Estimates} \\
& \multirow{2}{*}{True FDP} & (Benjamini \& & (Storey \& &
\multirow{2}{*}{(Efron, 2007)} \\ 
& &  Hochberg, 1995) &   Tibshirani, 2003) &  \\
\hline
``Cardio'' \hspace{10mm}  &&&&& \\
\multirow{2}{*}{\hspace{6mm} 40 tests} & 0.015 (0.01) & 0.0794 (0.013) &  0.0803 (0.015) & 0.0349 (0.016) \\
 & {\bf 0.00833 (0.0083)} &  {\bf 0.0311 (0.011)} &  {\bf 0.0301
   (0.011)} & {\bf 0.0469 (0.021)} \\
\cline{2-5}
\multirow{2}{*}{\hspace{6mm} 45 tests} & 0.0333 (0.014) & 0.144 (0.019) &  0.146 (0.021) & 0.0545 (0.021) \\
 & {\bf 0.0247 (0.013)} &  {\bf 0.0615 (0.02)} &  {\bf 0.0597 (0.019)} &  {\bf 0.0757 (0.031)} \\
\cline{2-5}
\multirow{2}{*}{\hspace{6mm} 50 tests} & 0.068 (0.017) & 0.294 (0.036) &  0.295 (0.037) & 0.0945 (0.024) \\
 &  {\bf 0.0578 (0.016)} & {\bf 0.106 (0.026)} &
   {\bf 0.104 (0.025)} &
  {\bf 0.1 (0.029)} \\
\cline{2-5}
\multirow{2}{*}{\hspace{6mm} 55 tests} & 0.131 (0.013) & 0.452 (0.068) &  0.453 (0.068) & 0.131 (0.025) \\
 &  {\bf 0.125 (0.015)} &  {\bf 0.196 (0.034)} &   {\bf 0.195 (0.033)} &  {\bf 0.153 (0.032)} \\
\cline{2-5}
\multirow{2}{*}{\hspace{6mm} 60 tests} & 0.19 (0.011) & 0.555 (0.072) &  0.555 (0.072) & 0.162 (0.022) \\
 &  {\bf 0.191 (0.013)} &  {\bf 0.26 (0.024)} &   {\bf 0.259 (0.023)} &  {\bf 0.178 (0.027)} \\
\hline
``Leukemia'' \hspace{6mm}  &&&&& \\
\multirow{2}{*}{\hspace{6mm} 40 tests} & 0.0875 (0.016) & 0.0777 (0.0068) &  0.073 (0.0064) & 0.288 (0.053) \\
 &  {\bf 0.0375 (0.012)} &  {\bf 0.0477 (0.013)} &   {\bf 0.0489 (0.013)} &  {\bf 0.0928 (0.026)} \\
\cline{2-5}
\multirow{2}{*}{\hspace{6mm} 45 tests} & 0.133 (0.023) & 0.119 (0.011) &  0.112 (0.012) & 0.324 (0.054) \\
 &  {\bf 0.0711 (0.016)} &  {\bf 0.0771 (0.017)} &   {\bf 0.08 (0.018)} &  {\bf 0.124 (0.031)} \\
\cline{2-5}
\multirow{2}{*}{\hspace{6mm} 50 tests} & 0.172 (0.021) & 0.162 (0.015) &  0.156 (0.015) & 0.36 (0.048) \\
 &  {\bf 0.122 (0.017)} &  {\bf 0.129 (0.023)} &   {\bf 0.132 (0.024)} &  {\bf 0.159 (0.033)} \\
\cline{2-5}
\multirow{2}{*}{\hspace{6mm} 55 tests} & 0.22 (0.019) & 0.194 (0.019) &  0.186 (0.02) & 0.383 (0.046) \\
 &  {\bf 0.178 (0.013)} &  {\bf 0.183 (0.021)} &   {\bf 0.186 (0.021)} &  {\bf 0.191 (0.03)} \\
\cline{2-5}
\multirow{2}{*}{\hspace{6mm} 60 tests} & 0.255 (0.015) & 0.243 (0.024) &  0.237 (0.024) & 0.405 (0.043) \\
 &  {\bf 0.223 (0.011)} &  {\bf 0.235 (0.022)} &   {\bf 0.24 (0.023)} &  {\bf 0.213 (0.03)} \\
\hline
\end{tabular}
} 
\end{center}
\caption{\em \footnotesize Results for simulations based on observed dependencies
  within the ``Cardio'' and ``Leukemia'' microarrays: True false
  discovery proportions (FDP) and FDR estimates with standard 
  errors are given when a 
  pre-specified number of tests are rejected.    Results using the
  sphering algorithm (in bold) are compared to data
  that has been row and array centered.  All data was simulated 
under the matrix decomposition model, \eqref{decomp}, with parameters
given in Section \ref{section_sim_mv}, and repeated ten times.  Two
sets of values should be compared: the true FDP with sheering to
without sphering, and the FDR estimates compared to the true FDP for
both with and without sphering.
\label{tab_res_micro}}
\end{table}

The results of the structured covariance study, namely improvements in
statistical power and FDR estimation, are confirmed on these
microarray-based simulations.  There are also some specific notes to
make regarding these simulations.  First in Table
\ref{tab_res_micro}, notice that using 
un-sphered data the FDR is overestimated on the 
``Cardio'' simulations and underestimated on the ``Leukemia''
simulations.  This is confirmed by the example giving the full FDR curves
in Figure \ref{fig_res_micro}.  After sphering, however, we see that
the FDR estimates for both simulations are still conservative, but
much closer to the true FDP.  
We note that all ten repetitions of the ``Cardio''
simulation estimated both $\hat{\Sig}$ and $\hat{\Delt}$ to be
non-diagonal.  This means that even after accounting for the correlations
among the genes, there still appear to be significant correlations
among the arrays.  In the ``Leukemia'' simulation, however,
$\hat{\Delt}$ was estimated to be diagonal in all ten simulations.
Thus, the correlations among the genes may be driving the
over-dispersion seen in the $t$-statistic distributions of Figure
\ref{fig_tstat_micro}.

Thus, from these simulations based on our matrix decomposition
model, we see that sphering the data as a pre-processing step
greatly improves statistical power and false discovery rate
estimation.

\subsection{Simulations: Other Models}
\label{section_sim_other}

We now evaluate the performance of our method using simulations based
on models other than the matrix-variate normal, namely a latent
variable model and a random effects model.  In these simulations we
will not only compare our sphering method to the standard method, but
also to the surrogate variable analysis method of \citet{leek_2008}.
We first compare this method and model's properties to our sphering
algorithm and matrix decomposition model, and then
compare these methods numerically.

To account for possible latent variables in a multiple testing
framework, \citet{leek_2008} propose a matrix model and the surrogate
variable analysis (SVA) method.  They propose the
model for the data $\X \in \Re^{m \times n}$, $\X = \mathbf{B} \Smat +
\mathbf{\Gamma} \mathbf{G} + 
\mathbf{U}$ where $\Smat$ is a signal matrix, $\mathbf{G} \in \Re^{d
  \times n}$ for $d < n$ is the
latent variable matrix, $\mathbf{U} \in \Re^{m \times n}$ is
independent noise and 
$\mathbf{B}$ and $\mathbf{\Gamma} \in \Re^{m \times d}$ are coefficients to be
estimated. 
 This model is similar in nature to our matrix
decomposition model \eqref{decomp}.  If we assume $\X$ has been
previously centered, using the notation of the latent variable model,
we can write \eqref{decomp} as $\X = \B \Smat + \Sig^{\frac{1}{2}}
\mathbf{U} \Delt^{\frac{1}{2}}$.  Thus, our model accounts for
structure within the data through the row and column covariances
$\Sig$ and $\Delt$, whereas their method estimates the structure through
$\mathbf{G}$ and assumes the noise is additive.  
Assuming that the latent variable model or our model is correct,
applying the respective algorithms results in 
approximately independent $p$-values.  Also similar to our method, SVA
can change the rankings of the test statistics.  Unlike SVA, however,
our model 
and sphering algorithm directly capture and account for possible
correlations among the columns as well as the rows.

\begin{table}[!!t]
\begin{center}
\scalebox{.8}{
\begin{tabular}{r|ll|ll|ll}
\hline
& \multicolumn{2}{|c|}{Standard} & \multicolumn{2}{|c|}{Sphered} &
\multicolumn{2}{|c}{SVA} \\
& FDP & $\widehat{\mathrm{FDR}}$ & FDP & $\widehat{\mathrm{FDR}}$ & FDP &
$\widehat{\mathrm{FDR}}$  \\
\hline
Latent Variable Model &&&&&& \\
\multirow{2}{*}{40 tests} &  0.08 & 0.168 &  0.05 & 0.0365 &  0.0528 & 0.0546 \\
 & (0.017) & (0.023) & (0.014) & (0.0086) & (0.01) & (0.0045) \\
\multirow{2}{*}{45 tests} &  0.109 & 0.212 &  0.0889 & 0.0657 &  0.0716 & 0.0868\\
 & (0.016) & (0.022) & (0.018) & (0.013) & (0.0097) & (0.0092)\\
\multirow{2}{*}{50 tests} &  0.15 & 0.303 &  0.124 & 0.106 &  0.127 & 0.118 \\
 & (0.018) & (0.035) & (0.017) & (0.021) & (0.01) & (0.015)\\
\multirow{2}{*}{55 tests} &  0.189 & 0.383 &  0.167 & 0.166 &  0.17 & 0.183\\
 & (0.015) & (0.051) & (0.018) & (0.021) & (0.012) & (0.02)\\
\multirow{2}{*}{60 tests} &  0.24 & 0.453 &  0.215 & 0.215 &  0.217 & 0.27\\
 & (0.011) & (0.05) & (0.017) & (0.023) & (0.0099) & (0.025)\\
\hline
Random Effects Model &&&&&& \\
\multirow{2}{*}{40 tests} &  0.375 & 0.011 &  0.0361 & 0.0318 &  0.4 & 0.136\\
 & (0.011) & (0.002) & (0.02) & (0.012) & (0.047) & (0.042) \\
\multirow{2}{*}{45 tests} &  0.433 & 0.0161 &  0.0642 & 0.0863 &  0.439 & 0.169\\
 & (0.0084) & (0.0031) & (0.026) & (0.031) & (0.042) & (0.048)\\
\multirow{2}{*}{50 tests} &  0.48 & 0.0196 &  0.102 & 0.14 &  0.475 & 0.185\\
 & (0.014) & (0.004) & (0.02) & (0.033) & (0.041) & (0.047)\\
\multirow{2}{*}{55 tests} &  0.52 & 0.0229 &  0.154 & 0.207 &  0.507 & 0.213\\
 & (0.013) & (0.0044) & (0.018) & (0.037) & (0.034) & (0.053)\\
\multirow{2}{*}{60 tests} &  0.553 & 0.0288 &  0.202 & 0.309 &  0.538 & 0.235\\
 & (0.012) & (0.0054) & (0.014) & (0.048) & (0.031) & (0.053)\\
\hline
\end{tabular}}
\end{center}
\caption{\em \footnotesize Simulations based on a latent variable and
  random-effects models as described in Section
  \ref{section_sim_other}.  The true FDP (FDP) is compared to the
  FDR estimates ($\widehat{\mathrm{FDR}}$) for a fixed number of rejected
  tests using the step-up method of \citet{bh_1995} for
  three pre-processing 
  techniques: Standard (row and column centered), our sphering
  algorithm, and the surrogate variable analysis (SVA) method.
  Averages are taken on ten repetitions and standard errors are given.
\label{tab_res_other}}
\end{table}

We simulate data from a latent variable model taken directly from
\citet{leek_2008} as well as from a random effects model denoting a
batch effect.  For both models, the data is of dimension $250 \times
50$, with 25 columns in each class with the following signal: The first 50
rows are non-null given by $\psi_{1, 1:25} =  0.5$, $\psi_{1,26:50} =
-0.5$, 
$\psi_{2,1:25} = -0.5$, $\psi_{2,26:50} = 0.5$ and the last 200
elements of $\psi_{1}$ and $\psi_{2}$ equal to zero.  For the latent
variable model, there are two latent variables given by $G_{ij}
\displaystyle\simiid^{iid} Bern(0.5)$, coefficients $\Gamma_{ij} \displaystyle\simiid^{iid}
N(0,1)$ and noise $U_{ij} \displaystyle\simiid^{idd} N(0,1)$.  For the
random-effects model with $K$ batches indicated by indices $I(k)$, a
column of the data is given 
by the following: 
$X_{rj} = \nu + \mu_{j} + \sum_{i=1}^{2} \psi_{i} I_{(j \in
  \mathcal{C}_{i})} + \sum_{k=1}^{K} \beta_{k} I_{(j \in I(k))} +
\epsilon$, where $\nu$, $\mu_{j}$ and $\psi_{i}$ are fixed effects,
and $\beta_{k} \displaystyle\simiid^{iid} N( \mu_{k}, \sigma^{2}_{k}
\mathbf{I})$ independent of
$\epsilon \displaystyle\simiid^{iid} N(0, \Sig_{1} )$  are random effects.
In our simulation, we 
have $\mu_{k} = [ -0.5 \hspace{2mm} -0.25 \hspace{2mm} 0 \hspace{2mm}
0.25 \hspace{2mm} 0.5 ]$, $\sigma^{2} = 0.5$, $\Sig = \Sig_{1}$ as
defined in Section \ref{section_sim_study} and $I(k)$ indicating
batches of five columns.

In Table \ref{tab_res_other}, we compare the true FDP to the estimate
of the FDR via the step-up method \citep{bh_1995} for the data with standard
pre-processing, with sphering and with \citet{leek_2008}'s surrogate
variable analysis (SVA).  The SVA method was implemented using the
defaults available in the package \textsf{sva} from \textsf{CRAN}, the
\textsf{R} language repository.
For the latent variable simulation, both our sphering method and the
SVA method improve the rank ordering of the test statistics resulting in higher
statistical power as well as improved estimates of the
FDR.  In the random-effects model simulation, however, the sphering
algorithm substantially outperforms the standard pre-processing 
and the SVA method.  We illustrate this by looking at the specific case
where 50 tests are rejected.  For the standard pre-processing and SVA methods,
the true FDP is 0.48 and 0.475 respectively, meaning that on average
25 out of the 50 
rejected tests are false positives.   With sphering, however, the order of
the test statistics is dramatically changed leading to a true FDP of
0.102 so that on average only 5 out of 50 rejected tests are
false.  The FDR estimates using the step-up method are also
problematic for the standard and SVA methods as 0.0196 and 0.185 are
substantially below the true FDPs of 0.48 and 0.475 respectively.  If
these methods were used, the number of false positives would not
be controlled.  With sphering, however, the FDR estimate of 0.14 is
much closer to the true FDP, 0.102 and is a conservative estimate, as
desired.

These simulations based on models other than the matrix-variate normal
reveal the robustness of our pre-processing technique.  Our sphering
method also compares very favorably to the surrogate variable analysis, another
pre-processing method.

\section{Discussion}
\label{section_discussion}

In this paper, we have demonstrated that using standard statistical
methodology to conduct inference on transposable data is problematic.
As a method of solving these problems, we have prosed a sphering
pre-processing technique that de-correlates the data yielding
approximately independent rows and columns.  We have revealed the
advantages and robustness of this method through simulations on many
correlated data sets.

The major disadvantage of our method is its computational cost.
Fitting the transposable regularized covariance model with $L_{1}$
penalties is approximately $O(k(m^{3} + p^{3}))$, where $k$ is the
total number of iterations needed until convergence.  Thus, directly
fitting this model to microarrays, for example, where $m$ may be
twenty or thirty 
thousand, is not currently feasible.  A simple fix can be proposed, that is to
first filter the genes by the absolute value of their un-sphered
$T$-statistics down to say 1,000 or 500 genes.  Since the signal in
each gene remains the same before and after sphering, filtering should
not effect the power to detect non-null genes, especially since
researchers are rarely interested in re-testing over 500 genes.  
As future work, we will examine approximations to the TRCM covariance
estimates that can be used in 
high-dimensional settings and would circumvent the need to filter the
genes before sphering.

There are many components of our work that deserve further
investigation and testing.  First, \citet{allen_2010} outline some of
the properties of the TRCM covariances estimates, but several
questions, such as the consistency of the estimates, remain.  Also,
direct estimation of $\eta$, the scaled variance of the $Z$-statistic
that depends on the array covariance, should be examined to find a
consistent estimate of $\eta$.

In conclusion, our model and study have revealed several important
issues related to large-scale inference with transposable data such as
microarrays.  First, correlation among the columns proves to be
a major problem, both theoretically and in simulations, when comparing
test statistics to a theoretical or permutation null distribution.
This results is striking as inference is often conducted under the false
assumption of column independence.   Second, despite
the lack of theoretical results supporting the use of many common
FDR-controlling procedures for test statistics with arbitrary dependence
structure, the procedures
seem to conservatively estimate the FDP under a variety of
correlation scenarios.   Finally, our method of de-correlating the
data is a way to directly model the covariance structure in a
multiple testing framework.  This method leads to 1) improvements in the
statistical power, and to
2) better estimation of the FDR.  
While this paper has focused on the
example of two-class microarrays, our model and methods may prove
useful in a variety large-scale inference problems with highly
transposable data sets.

\section{Acknowledgments}

We would like to thank Jonathan Taylor for several helpful comments and
conversations regarding this work.  Thanks to Joseph Romano for
discussions and references for papers on multiple testing with dependencies.
Thanks also to Bradley Efron whose
observations and ideas on microarrays partly inspired this work.

\appendix

\section{Additional Simulation Results}
\label{section_app_sims}

\begin{table}
\begin{center}
\scalebox{.8}{
\begin{tabular}{r|l|llll}
\hline
& & \multicolumn{3}{c}{FDR Estimates} \\
& \multirow{2}{*}{True FDP} & (Benjamini \& & (Storey \& & 
\multirow{2}{*}{(Efron, 2007)} \\ 
& &  Hochberg, 1995) &   Tibshirani, 2003) &   \\
\hline
$\Sig_{1}, \Delt_{2}$ \hspace{10mm}  &&&& \\
\multirow{2}{*}{\hspace{6mm} 40 tests} & 0.0125 (0.0077) & 0.0831 (0.018) &  0.0783 (0.019) & 0.0749 (0.024) \\
 & {\bf 0.0194 (0.013)} & {\bf 0.022 (0.0074)} &  {\bf 0.0215 (0.0072)} & {\bf 0.0495 (0.029)} \\
\cline{2-5}
\multirow{2}{*}{\hspace{6mm} 45 tests} & 0.0333 (0.015) & 0.164 (0.031) &  0.16 (0.032) & 0.117 (0.032) \\
 & {\bf 0.0321 (0.018)} & {\bf 0.0436 (0.011)} &  {\bf 0.0431 (0.011)} & {\bf 0.0761 (0.03)} \\
\cline{2-5}
\multirow{2}{*}{\hspace{6mm} 50 tests} & 0.078 (0.015) & 0.281 (0.027) &  0.276 (0.028) & 0.165 (0.032) \\
 & {\bf 0.0556 (0.02)} & {\bf 0.0883 (0.013)} &  {\bf 0.0875 (0.013)} & {\bf 0.117 (0.031)} \\
\cline{2-5}
\multirow{2}{*}{\hspace{6mm} 55 tests} & 0.135 (0.012) & 0.368 (0.028) &  0.364 (0.028) & 0.194 (0.028) \\
 & {\bf 0.123 (0.015)} & {\bf 0.181 (0.02)} &  {\bf 0.182 (0.02)} & {\bf 0.179 (0.031)} \\
\cline{2-5}
\multirow{2}{*}{\hspace{6mm} 60 tests} & 0.198 (0.011) & 0.461 (0.044) &  0.458 (0.045) & 0.234 (0.028) \\
 & {\bf 0.194 (0.014)} & {\bf 0.242 (0.023)} &  {\bf 0.244 (0.023)} & {\bf 0.201 (0.033)} \\
\hline
$\Sig_{2}, \Delt = \mathbf{I}$ \hspace{6mm}  &&&& \\
\multirow{2}{*}{\hspace{6mm} 40 tests} & 0.035 (0.017) & 0.0498 (0.0081) &  0.0513 (0.0082) & 0.161 (0.034) \\
 & {\bf 0.03 (0.015)} & {\bf 0.0279 (0.0058)} &  {\bf 0.0279 (0.0055)} & {\bf 0.081 (0.029)} \\
\cline{2-5}
\multirow{2}{*}{\hspace{6mm} 45 tests} & 0.0711 (0.019) & 0.0839 (0.012) &  0.0856 (0.012) & 0.209 (0.041) \\
 & {\bf 0.0644 (0.017)} & {\bf 0.0607 (0.0096)} &  {\bf 0.0604 (0.0094)} & {\bf 0.133 (0.035)} \\
\cline{2-5}
\multirow{2}{*}{\hspace{6mm} 50 tests} & 0.12 (0.018) & 0.141 (0.021) &  0.143 (0.021) & 0.249 (0.045) \\
 & {\bf 0.098 (0.013)} & {\bf 0.0989 (0.013)} &  {\bf 0.0985 (0.013)} & {\bf 0.161 (0.035)} \\
\cline{2-5}
\multirow{2}{*}{\hspace{6mm} 55 tests} & 0.167 (0.016) & 0.191 (0.029) &  0.192 (0.029) & 0.278 (0.04) \\
 & {\bf 0.14 (0.009)} & {\bf 0.158 (0.016)} &  {\bf 0.16 (0.016)} & {\bf 0.208 (0.035)} \\
\cline{2-5}
\multirow{2}{*}{\hspace{6mm} 60 tests} & 0.217 (0.014) & 0.243 (0.035) &  0.246 (0.035) & 0.311 (0.041) \\
 & {\bf 0.197 (0.0065)} & {\bf 0.227 (0.017)} &  {\bf 0.229 (0.017)} & {\bf 0.242 (0.034)} \\
\hline
$\Sig_{2}, \Delt_{1}$ \hspace{10mm}  &&&& \\
\multirow{2}{*}{\hspace{6mm} 40 tests} & 0.005 (0.005) & 0.0455 (0.0065) &  0.0439 (0.0065) & 0.00846 (0.0041) \\
 & {\bf 0 (0)} & {\bf 0.00305 (0.0014)} &
  {\bf 0.00267 (0.0012)} &
 {\bf 0.00185 (0.001)} \\
\cline{2-5}
\multirow{2}{*}{\hspace{6mm} 45 tests} & 0.0111 (0.005) & 0.111 (0.027) &  0.11 (0.026) & 0.0198 (0.0076) \\
 & {\bf 0 (0)} & {\bf 0.00845 (0.0031)} &  {\bf 0.00783 (0.0029)} & {\bf 0.00656 (0.0026)} \\
\cline{2-5}
\multirow{2}{*}{\hspace{6mm} 50 tests} & 0.042 (0.0081) & 0.225 (0.046) &  0.225 (0.047) & 0.0493 (0.014) \\
 & {\bf 0.03 (0.0061)} & {\bf 0.0436 (0.0076)} &  {\bf 0.0433 (0.0075)} & {\bf 0.033 (0.0083)} \\
\cline{2-5}
\multirow{2}{*}{\hspace{6mm} 55 tests} & 0.109 (0.0086) & 0.404 (0.034) &  0.409 (0.034) & 0.0923 (0.017) \\
 & {\bf 0.0964 (0.0039)} & {\bf 0.118 (0.014)} &  {\bf 0.118 (0.015)} & {\bf 0.0756 (0.014)} \\
\cline{2-5}
\multirow{2}{*}{\hspace{6mm} 60 tests} & 0.178 (0.0056) & 0.552 (0.048) &  0.554 (0.048) & 0.133 (0.018) \\
 & {\bf 0.168 (0.0017)} & {\bf 0.214 (0.014)} &  {\bf 0.216 (0.014)} & {\bf 0.114 (0.015)} \\
\hline
$\Sig_{2}, \Delt_{2}$ \hspace{10mm}  &&&& \\
\multirow{2}{*}{\hspace{6mm} 40 tests} & 0.0075 (0.0053) & 0.0712 (0.016) &  0.066 (0.015) & 0.0444 (0.012) \\
 & {\bf 0.0125 (0.0077)} & {\bf 0.0108 (0.0021)} &  {\bf 0.0104 (0.0019)} & {\bf 0.0152 (0.0042)} \\
\cline{2-5}
\multirow{2}{*}{\hspace{6mm} 45 tests} & 0.0311 (0.011) & 0.169 (0.022) &  0.163 (0.021) & 0.087 (0.019) \\
 & {\bf 0.0244 (0.0084)} & {\bf 0.0317 (0.0056)} &  {\bf 0.0309 (0.0057)} & {\bf 0.0336 (0.007)} \\
\cline{2-5}
\multirow{2}{*}{\hspace{6mm} 50 tests} & 0.078 (0.012) & 0.277 (0.025) &  0.271 (0.025) & 0.132 (0.021) \\
 & {\bf 0.072 (0.015)} & {\bf 0.092 (0.016)} &  {\bf 0.0918 (0.016)} & {\bf 0.0738 (0.011)} \\
\cline{2-5}
\multirow{2}{*}{\hspace{6mm} 55 tests} & 0.144 (0.013) & 0.388 (0.037) &  0.383 (0.037) & 0.167 (0.019) \\
 & {\bf 0.136 (0.013)} & {\bf 0.162 (0.013)} &  {\bf 0.162 (0.013)} & {\bf 0.116 (0.013)} \\
\cline{2-5}
\multirow{2}{*}{\hspace{6mm} 60 tests} & 0.198 (0.013) & 0.47 (0.044) &  0.468 (0.043) & 0.197 (0.02) \\
 & {\bf 0.202 (0.012)} & {\bf 0.223 (0.014)} &  {\bf 0.223 (0.014)} & {\bf 0.143 (0.01)} \\
\hline
\end{tabular}
}
\end{center}
\caption{\em \footnotesize Additional simulation study results: True false
  discovery proportions (FDP) and FDR estimates with standard 
  errors are given when a 
  pre-specified number of tests are rejected.    Results using the 
  sphering algorithm (in bold) are compared to data
  that has been row and column centered.  All data was simulated 
under the matrix decomposition model, \eqref{decomp}, with parameters
given in Section \ref{section_sim_study}, and repeated ten times. Two
sets of values should be compared: the true FDP with sphering to
without sphering, and the FDR estimates compared to the true FDP for
both with and without sphering. }
\end{table}

\section{Proofs}
\label{section_proofs}

\begin{proof}[Theorem \ref{prop1}]
{\footnotesize
Let $\mathbf{z}$ be a vector of $N(0,1)$ random variables.  Then, if
we arrange $\x$ as a column vector, we have
\begin{align*}
\x_{(n)} \eqdist^{d} \left( \begin{array}{c} \psi_{1} \mathbf{1}_{(n_{1})}
     \\ \psi_{2} \mathbf{1}_{(n_{2})} \end{array} \right) +
\sigma \mathbf{L} \mathbf{z}_{(n)}.
\end{align*}
Thus, we can write $Z$ as a sum of the scaled independent and normally
distributed random variables $\mathbf{z}$.  The expected value of $Z$
is trivial and the variance can be written as the following.
\begin{align*}
\mathrm{Var}(Z) &= \frac{1}{\sigma^{2} c_{n}} \mathrm{Var} \left(
  \bar{\x}_{1} - \bar{\x}_{2} \right) \\
&= \frac{1}{c_{n}} \mathrm{Var}\left( \frac{1}{n_{1}}
  \sum_{i \in \mathcal{C}_{1}} ( \mathbf{L} \mathbf{z} )_{i} -
  \frac{1}{n_{2}} \sum_{i \in \mathcal{C}_{2}} ( \mathbf{L} \mathbf{z}
  )_{i}  \right) \\
&= \frac{1}{c_{n}} \mathrm{Var} \left[ \sum_{j=1}^{n} \left( \frac{1}{n_{1}}
  \sum_{i \in \mathcal{C}_{1}} L_{ij} z_{j}  - \frac{1}{n_{2}}
  \sum_{i \in \mathcal{C}_{2}} L_{ij} z_{j} \right)  \right] \\
&= \frac{\eta}{c_{n}}
\end{align*} 
The last step follows since the $z_{j}$'s are independent.
Note also that if we let $W_{i} \triangleq \begin{cases}
  \frac{1}{n_{1}} & i \in \mathcal{C}_{1} \\ - \frac{1}{n_{2}} & i \in
  \mathcal{C}_{2} \end{cases}$, we can write $\eta = \sum_{i=1}^{n}
\sum_{j=1}^{n} \Delt_{ij} W_{i} W_{j}$.  }
\end{proof}

\begin{proof}[Corollary \ref{cor_1}]
{\footnotesize
This is trivial following the proof of Theorem \ref{prop1} since
the matrix square root of $\Delt$ can be written as
$\mathbf{L} = \left( \begin{array}{cc} \mathbf{L}_{1} & \mathbf{0}
    \\ \mathbf{0} & \mathbf{L}_{2} \end{array} \right)$. }
\end{proof}

\begin{proof}[Proposition \ref{sphered_dist}]
{\footnotesize
For part (i), the matrix decomposition model implies that
$\mathrm{E}(X_{ij}) = \nu_{i} + \mu_{j} + \psi_{k,i}$ for $k=1,2$
depending on the class of array $j$.  Each element of the noise as
defined in Step 2 can be written as $N_{ij} = X_{ij} - \hat{\mu}_{j} -
\hat{\nu}_{i} - \hat{\psi}_{k,i}$ where $\hat{\mu}_{j} = \frac{1}{n}
\sum_{i=1}^{n} X_{ij}$, $\hat{\nu}_{i} = \frac{1}{p} \sum_{j=1}^{p}
(X_{ij} - \hat{\mu}_{j})$, and $\hat{\psi}_{k,i} = \frac{1}{n_{k}}
\sum_{j \in \mathcal{C}_{k}} ( X_{ij} - \hat{\mu}_{j} -
\hat{\nu}_{i})$.  We show that $\mathrm{E}(N_{ij}) = 0$, which in the
process proves that $\mathrm{E}(\tilde{X}_{ij}) = \psi_{k,i}$.
\begin{align*}
\mathrm{E}(\hat{\mu}_{j} + \hat{\nu}_{i} + \hat{\psi}_{k,i}) &=
\frac{1}{n} \sum_{i=1}^{n} \mathrm{E}(X_{ij}) + \frac{1}{n_{k}}
 \sum_{j \in \mathcal{C}_{k}} \mathrm{E}(X_{ij}) - \frac{1}{n}
 \frac{1}{n_{k}} \sum_{j \in \mathcal{C}_{k}} \sum_{i=1}^{n}
 \mathrm{E}(X_{ij}) \\
&= \mu_{j} + \bar{\nu} + \bar{\psi}_{k} + \bar{\mu}_{k} + \nu_{i} -
\psi_{k,i}  - \bar{\mu}_{k} - \bar{\nu} - \bar{\psi}_{k} \\
&= \mu_{j} + \nu_{i} + \psi_{k,i}.
\end{align*}

Now for part (ii), $\tilde{\X} - \Smat = \tilde{\N} \triangleq
\hat{\Sig}^{-\frac{1}{2}} \N \hat{\Delt}^{-\frac{1}{2}}$, and
following the proof of part (i), $\N \sim N_{m,n}( \mathbf{0},
\mathbf{0}, \Sig, \Delt)$. The characteristic function of the centered
matrix-variate normal is $\phi_{\X}(\mathbf{Z}) = \mathrm{etr}(
-\frac{1}{2} \Z^{T} \Sig \Z \Delt)$ where $\mathrm{etr}$ is the
exponential of the trace function \citep{gupta_nagar}.  The
characteristic function of $\tilde{\N}$ can then be written as
\begin{align*}
\phi_{\tilde{\N}}(\Z) &= \mathrm{etr} \left[ -\frac{1}{2} \left(
    \hat{\Sig}^{-\frac{1}{2}} \Z \hat{\Delt}^{-\frac{1}{2}}
  \right)^{T} \Sig \left(
    \hat{\Sig}^{-\frac{1}{2}} \Z \hat{\Delt}^{-\frac{1}{2}}
  \right) \Delt \right] \\
&= \mathrm{etr} \left[ -\frac{1}{2} 
    \hat{\Delt}^{-\frac{1}{2}} \Z^{T} \hat{\Sig}^{-\frac{1}{2}} \Sig
    \hat{\Sig}^{-\frac{1}{2}} \Z \hat{\Delt}^{-\frac{1}{2}}  \Delt
  \right] \\
&= \mathrm{etr} \left[ -\frac{1}{2} 
     \Z^{T} \left( \hat{\Sig}^{-\frac{1}{2}} \Sig
    \hat{\Sig}^{-\frac{1}{2}} \right) \Z \left( \hat{\Delt}^{-\frac{1}{2}}  \Delt
    \hat{\Delt}^{-\frac{1}{2}} \right)  \right].
\end{align*}
Thus, letting $\tilde{\Sig} = \hat{\Sig}^{-\frac{1}{2}} \Sig
\hat{\Sig}^{-\frac{1}{2}}$ and $\tilde{\Delt} = \hat{\Delt}^{-\frac{1}{2}} \Delt
\hat{\Delt}^{-\frac{1}{2}}$, we have $\tilde{\N} \sim N_{m,n}(\mathbf{0},
\mathbf{0}, \tilde{\Sig}, \tilde{\Delt})$. 

}
\end{proof}

\begin{proof}[Proposition \ref{sphering_t}]
{\footnotesize
Since we are considering the test statistic for one gene, we will
suppress the index $i$.  We can define the random variables $Z
\triangleq ( \bar{X}_{1} - \bar{X}_{2}) / \sigma \sqrt{c_{n}}$ and
$D \triangleq s^{2}_{\tilde{X}_{1} \tilde{X}_{2}} / \sigma^{2}$.  Then,
$\tilde{T}$ can be written as $\tilde{T} = Z / \sqrt{D}$.  From
Theorem \ref{prop1}, $Z \sim N( (\psi_{1} - \psi_{2})/\sigma
  \sqrt{c_{n}}, \eta / c_{n})$.  Then, under the null, $Z /
\sqrt{ \eta / c_{n} } \sim N(0, 1)$.  Also, under the null, $D
 \sigma^{2} / \tilde{\sigma}^{2}  \sim \chi^{2}_{(n-2)}$ with $D$
and $Z$ independent.  Then,
\begin{align*}
\frac{Z / \sqrt{\frac{\eta}{c_{n}}}}{\sqrt{D
    \frac{\sigma^{2}}{\tilde{\sigma}^{2}}}} \sim t_{(n-2)}
\hspace{2mm} \Rightarrow \hspace{2mm} \tilde{T} = \frac{Z}{\sqrt{D}}
\sim \frac{\tilde{\sigma}}{\sigma} \sqrt{\frac{\eta}{c_{n}}} t_{(n-2)}.
\end{align*}
}
\end{proof}

\bibliographystyle{Chicago}
\bibliography{trcm}

\end{document}